\documentclass[12pt]{article}
\usepackage{jheppub}
\usepackage{subcaption}

\newcommand{\cc}{{\cal C}}

\newcommand{\cn}{{\cal N}}

\newcommand{\cl}{{\cal L}}
\newcommand{\cf}{{\cal F}}

\newcommand{\cg}{{\cal G}}

\newcommand{\co}{{\cal O}}

\def\bal#1\eal{\begin{align}#1\end{align}}
\def\alp[#1]{\begin{align}#1\end{align}}

\def\secnum[#1]{\texorpdfstring{$#1$}{TEXT}}

\def\secnuml#1\secnumr{\texorpdfstring{$#1$}{TEXT}}

\def\eqa{\begin{eqnarray}}
\def\eqae{\end{eqnarray}}
\def\eq{\begin{equation}}
\def\eqe{\end{equation}}
\def\be{\begin{equation}}
\def\ee{\end{equation}}
\def\bea{\begin{eqnarray}}
\def\eea{\end{eqnarray}}
\def\ba{\begin{array}}
\def\ea{\end{array}}
\def\bd{\begin{displaymath}}
\def\ed{\end{displaymath}}

\def\>{\rangle}
\def\<{\langle}
\def\a{\alpha}
\def\b{\beta}
\def\c{\chi}
\def\del{\delta}

\def\f{\phi}

\def\g{\gamma}

\def\l{\lambda}

\def\n{\nu}

\def\p{\pi}

\def\t{\tau}

\def\F{\Phi}

\def\S{\Sigma}

\def\pa{\partial}

\newcommand{\bz}{{\mathbb{Z}}}

\title{\Large Correlators in the $\cn=2$ Supersymmetric SYK Model}

\author{Cheng Peng, Marcus Spradlin and Anastasia Volovich}
\affiliation{Department of Physics, Brown University, Providence RI 02912, USA}
\emailAdd{cheng$\underline{~}$peng@brown.edu, marcus$\underline{~}$spradlin@brown.edu, anastasia$\underline{~}$volovich@brown.edu}

\abstract{We study correlation functions in the one-dimensional $\mathcal{N}=2$ supersymmetric SYK model. The leading order 4-point correlation functions are computed by summing over ladder diagrams expanded in a suitable basis of conformal eigenfunctions. A novelty of the $\cn=2$ model is that both symmetric and antisymmetric eigenfunctions are required. Although we use a component formalism, we verify that the operator spectrum and 4-point functions are consistent with $\mathcal{N}=2$ supersymmetry. We also confirm the maximally chaotic behavior of this model and  comment briefly on its 6-point functions.}

\begin{document}

\maketitle

\section{Introduction}

Extensive studies of the Sachdev-Ye-Kitaev (SYK)  model~\cite{Sachdev:1992fk,PG,GPS,KitaevTalk1,KitaevTalk2,Maldacena:2016hyu} have revealed several fascinating features of this solvable large-$N$ model. Perhaps the most important property is its quantum chaotic behavior~\cite{KitaevTalk1,KitaevTalk2,Maldacena:2016hyu} that makes it a promising, but still somewhat mysterious, candidate for a holographic dual of $AdS_2$ quantum gravity~\cite{Strominger:1998yg,Maldacena:1998uz,Almheiri:2014cka,Shenker:2013pqa,Shenker:2014cwa,Maldacena:2015waa,Sachdev:2015efa,Sachdev:2010um, Maldacena:2017axo,Blake:2016jnn}.  The model develops an emergent (approximate) reparametrization symmetry at low energy~\cite{Maldacena:2016hyu,Bagrets:2016cdf,Stanford:2017thb,Mertens:2017mtv} that is also present in dilaton gravity theories on AdS$_2$~\cite{Maldacena:2016hyu,Maldacena:2016upp,Engelsoy:2016xyb, Cvetic:2016eiv,Forste:2017kwy}. It has intimate relations with well-studied random matrix models~\cite{Polchinski:2016xgd,Jevicki:2016bwu,Maldacena:2016hyu,Jevicki:2016ito,Gross:2017hcz, Garcia-Garcia:2016mno, Garcia-Garcia:2017pzl, Cotler:2016fpe,Liu:2016rdi,Stanford:2017thb,Li:2017hdt}, it further boosts the study of a different type of the large-$N$ limit~\cite{Witten:2016iux,Gurau:2011xq,Bonzom:2011zz, Bonzom:2012hw,Gurau:2016lzk,Klebanov:2016xxf,Carrozza:2015adg,Nishinaka:2016nxg,Peng:2016mxj, Krishnan:2016bvg,Krishnan:2017ztz, Gurau:2017xhf,Bonzom:2017pqs,Narayan:2017qtw,Klebanov:2017nlk,Mironov:2017aqv}, and it is closely related to vector models~\cite{Peng:2017kro}. The SYK model can be generalized to include extra symmetries~\cite{Gross:2016kjj} or to live in higher dimensions~\cite{Gu:2016oyy,Berkooz:2016cvq,Jian:2017unn,Turiaci:2017zwd,Gu:2017ohj,Das:2017pif}. Of course, the model is also of great interest in condensed matter physics~\cite{Danshita:2016xbo,Garcia-Alvarez:2016wem,
Davison:2016ngz,You:2016ldz,Fu:2016yrv,Hartnoll:2016apf,Gu:2016oyy, Banerjee:2016ncu,Bi:2017yvx,Jian:2017jfl,Gu:2017ohj, Wu:2017exh,Song:2017pfw}. Other related recent work can be found in~\cite{Jensen:2016pah,Ferrari:2017ryl, Berkooz:2017efq,Bagrets:2017pwq,Caputa:2017urj,Chowdhury:2017jzb,Cotler:2017abq,Itoyama:2017emp,Ho:2017nyc,Itoyama:2017xid,Gurau:2017qna,Dartois:2017xoe,Blake:2017qgd,Gurau:2017qya,Krishnan:2017txw, Kanazawa:2017dpd}.

In particular, supersymmetric extensions of the SYK model have been constructed in~\cite{Fu:2016vas}, and
a supersymmetric SYK-like model without a random coupling has been proposed in~\cite{Peng:2016mxj}.
Aspects of supersymmetric SYK models have also been studied in~\cite{Gross:2016kjj, Anninos:2016szt, Sannomiya:2016mnj, Li:2017hdt}.
In this paper, we study the correlation functions of the $\cn=2$ supersymmetric SYK model proposed in~\cite{Fu:2016vas}. One nice feature of the $\cn=2$ model, compared to its $\cn=1$ cousin, is that the supersymmetry is preserved not only in the large-$N$ limit, but also for finite $N$~\cite{Fu:2016vas}.

In section 2 we review
the $\cn=2$ model, which has a few new technical features compared to the $\cn=1$ theory. In section 3 we discuss the fact that in contrast to both
the fermionic SYK model studied
in~\cite{KitaevTalk1,KitaevTalk2,Maldacena:2016hyu} and the
$\cn = 1$ supersymmetric model studied
in~\cite{Fu:2016vas},
the computation of 4-point functions in
the $\cn=2$ model requires both antisymmetric and symmetric
conformal eigenfunctions, which we work out.
In section 4 we compute the 4-point functions and see another difference:
 divergences in 4-point functions arise from the exchange of a full $\cn=2$ supermultiplet.
This $\cn=2$ supermultiplet presumably leads to a super-Schwarzian effective action due to the breaking of conformal symmetry in the infared. While these new features were anticipated~\cite{Fu:2016vas}, we provide some concrete computations to confirm them.
Finally in section 5 we briefly comment on the operator product expansion (OPE) between operators that are  bilinear in the fundamental fields.
These OPE coefficients may be extracted from our 4-point function results
in a manner almost identical to the analysis of~\cite{Gross:2017hcz}.

\bigskip
\noindent
{\bf Note Added}

As we were preparing this work for submission we became aware of
other work~\cite{Murugan:2017eto,Yoon:2017gut} on supersymmetric SYK models.
Among many other things,
the paper~\cite{Murugan:2017eto} studies the correlation functions of
the $\cn=1$ supersymmetric SYK model using the formalism of a real
superfield.
In this work we
focus on the $\cn=2$ supersymmetric SYK model which could be constructed
using a complex superfield, although we use component fields.
Our paper therefore also overlaps with the newly appearing paper~\cite{Bulycheva:2017uqj} on the SYK model
with complex fermions.

\section{The Operator Spectrum}

In this paper we study the $\cn=2$ supersymmetric SYK model of~\cite{Fu:2016vas}. We begin in this section by supplying some details about the spectrum of this model that were not given explicitly in~\cite{Fu:2016vas}.
The model describes $N$ complex fermions $\psi_i$ in $1{+}0$ dimensions governed by the Lagrangian
\bal
\cl&=i \bar{\psi}_i\pa\psi_i-\bar{b}_ib_i+  i^{\frac{q-1}{2}} C_{i{j_1}\ldots {j_{q-1}}}\bar{b}_i\psi_{j_1}\ldots\psi_{j_{q-1}}+ i^{\frac{q-1}{2}} \bar{C}_{i{j_1}\ldots {j_{q-1}}}{b}_i\bar{\psi}_{j_1}\ldots\bar{\psi}_{j_{q-1}}\,,
\eal
where $b$ is a complex bosonic auxiliary field, $q$ is an odd integer (so that the Lagrangian is bosonic), and $C_{i_1\cdots i_q}$ is a random complex coupling drawn from a Gaussian distribution with
\bal
\<C_{i_i\ldots i_q}\bar{C}_{i_i\ldots i_q}\>=\frac{(q-1)!\, J}{N^{ q-1 }}\,.\label{norm}
\eal
We are interested in the large-$N$ limit, with $J$ held fixed.

\subsection{2-point functions}

We begin by considering the 2-point functions
\begin{equation}
G^\psi(\t_{12}) = \frac{1}{N} \sum_{i=1}^N \langle \psi_i(\t_1) \bar{\psi}_i(\t_2) \rangle\,,
\qquad
G^b(\t_{12}) = \frac{1}{N} \sum_{i=1}^N \langle b_i(\t_1) \bar{b}_i(\t_2) \rangle
\end{equation}
in Euclidean time $\t$, where we use $\t_{ij} = \t_i - \t_j$.
The complex conjugate 2-point functions, obtained by replacing $\psi \leftrightarrow \bar{\psi}$
and $b \leftrightarrow b$, evidently satisfy
\begin{equation}
G^{\bar\psi }(\t)=-G^{\psi }(-\t)\,,\qquad  G^{\bar b }(\t)=G^{b }(-\t)\,.\label{conjpp}
\end{equation}
In the IR limit, the Schwinger-Dyson equations relating the 2-point functions
to the corresponding
self-energies read
\bal
\S^\psi(\t_{12})&=(q-1)J  G^{b}(\t_{12})(G^{\bar\psi}(\t_{12}))^{q-2}\,,\\
\S^b(\t_{12})&=J(G^{\psi}(\t_{12}))^{q-1}\,,\\
- \delta(\t_{13}) &= \int d\t_2 G^\psi(\t_{12}) \S^\psi(\t_{23})\,,\\
-\delta(\t_{13}) &= \int d\t_2 G^b (\t_{12}) \S^b(\t_{23})\,,
\eal
together with their complex conjugates.
Taking the ansatz
\begin{equation}
G^{b}(\t_{12})= \frac{b_b}{|\t_{12}|^{2\Delta_{b}}}\,,\qquad
G^{\psi}(\t_{12})= \frac{b_\psi \,\text{sgn}(\t_{12})}{|\t_{12}|^{2\Delta_{\psi}}}\,,
\label{res1}
\end{equation}
the solution is found to be~\cite{Fu:2016vas}
\begin{equation}
\Delta_\psi=\frac{1}{2q}\,,\quad \Delta_b=\frac{1+q}{2q}\,,\quad
  b_\psi=\Big(\frac{\tan\frac{\p}{2q}}{2\p J}\Big)^{1/q}\,,\quad b_b=\frac{1}{q}\Big(\frac{\tan\frac{\p}{2q}}{2\p J}\Big)^{1/q}\,.
\label{res2}
\end{equation}
From these results and the relation~\eqref{conjpp} we see that
\begin{equation}
 G^{\psi}(\t)=G^{\bar{\psi}}(\t)\,, \qquad  G^{b}(\t)=G^{\bar{b}}(\t)\,.\label{conjres}
\end{equation}

\subsection{The diagonal 4-point kernel}

\begin{figure}[t]
\centering
\includegraphics[width=0.20\linewidth]{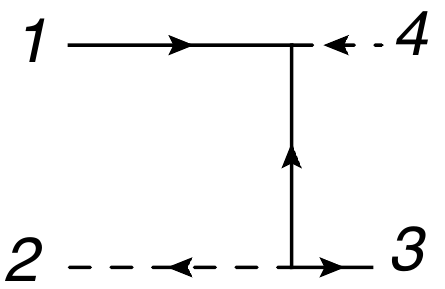}
\caption{Kernel of the $\<\psi_i(\t_1)b_i(\t_2)\bar{\psi}_j(\t_3)\bar{b}_j(\t_4)\>$ correlation. Iterating this kernel generates all ladder diagrams, which dominate the large-$N$ limit  of the connected 4-point function.}
\label{fig:sk1}
\end{figure}

In the large-$N$ limit, the connected 4-point functions are dominated by ladder
diagrams. This will be the topic of section~\ref{fourpointsection}; here we only need
to recall that these can be generated iteratively by repeated convolution with
an appropriate integral kernel.
The kernels of the different types of 4-point functions can be worked out straightforwardly.
For $\<\psi_i(\t_1)b_i(\t_2)\bar{\psi}_j(\t_3)\bar{b}_j(\t_4)\>$ the kernel is (see figure~\ref{fig:sk1})
\bal
K^{d}
=J (q-1) G^\psi(\t_{14})G^b(\t_{23})(G^{\psi}(\t_{34}))^{q-2}\,,\label{kergen}
\eal
where the factor of $q-1$ arises from the $(q-1)!$ in~\eqref{norm} divided by a symmetry factor $(q-2)!$.
The superscript ``$d$'' indicates that this kernel is diagonal
in the sense that
the directions of the arrows on the left and right sides of the kernel match.  This means
it can be iterated directly to build ladder diagrams with arbitrarily many rungs.
The operators running in the OPE channel of this kernel take the schematic form $\psi_i\pa^n b_i$, and have $U(1)$ charge $\frac{1}{q}+\frac{q-1}{q}=1$.

There is another kernel which is almost the same as this, but with all fields replaced by their conjugates. We could call this $K^{\bar{d}}$, but it follows from eq.~\eqref{conjres} that $K^{\bar{d}} = K^d$. This simply means that there is another set of conjugate operators $\bar{\psi}_i \partial^n \bar{b}_i$ with the same dimensions as the ones corresponding to the kernel $K^d$.

In the conformal limit the kernel becomes simply
\begin{equation}
\label{conformalKd}
K_c^{d}(\t_1, \t_2; \t_3, \t_4)=
 \frac{\a\,\text{sgn}(\t_{14})\text{sgn}(\t_{34})}{|\t_{14}|^{1/q}|\t_{34}|^{(q-2)/q}|\t_{23}|^{(1+q)/q}}\,,
\end{equation}
where
\begin{equation}
\a = \frac{q-1}{q} \frac{\tan(\frac{\pi}{2q})}{2 \pi}\,.
\end{equation}
Next we consider the kernel convolution eigen-equation
\begin{equation}
k\,f(\t_1,\t_2)=
\int d\tau_3 d\tau_4 K(\t_1,\t_2;\t_3,\t_4) f(\t_3,\t_4)\,,
\end{equation}
which for a generic kernel
admits both symmetric and antisymmetric eigenfunctions $f(\t_1,\t_2)$.
We will denote the symmetric and antisymmetric eigenvalues of the kernel~\eqref{conformalKd}
by $k_c^{s,d}$ and $k_c^{a,d}$, respectively.
Here, and in all that follows, the superscripts ``$s$'' and ``$a$'' stand respectively for symmetric
and antisymmetric, the subscript ``$c$'' reminds us that we are working in the conformal limit,
and the superscript ``$d$'' indicates that these are the eigenvalues of the diagonal kernel $K^d$.

As described in section 3.2.3 of~\cite{Maldacena:2016hyu}, conformal invariance
effectively allows the eigenvalues to be determined simply by
\bal
k_c^{s,d} &= \int d\t_3d\t_4 K_c^{d}(1,0;\t_3,\t_4) \frac{1}{|\t_{34}|^{\Delta_\psi+\Delta_b-h}}\,, \label{ees}\\
k_c^{a,d} &=  \int d\t_3d\t_4 K_c^{d}(1,0;\t_3,\t_4) \frac{\text{sgn}(\t_{34})}{|\t_{34}|^{\Delta_\psi+\Delta_b-h}} \label{eea}
\eal
in terms of a conformal weight $h$.
Note that in this case the two outgoing lines are a boson and a fermion, so we use $\Delta_\psi+\Delta_b-h$ in the eigenfunction instead of $2 \Delta_\psi-h$ as in~\cite{Maldacena:2016hyu}.
Plugging in eqs.~\eqref{res1} and~\eqref{res2}, the eigen-equations become
\bal
k_c^{s,d} &=\a \int d\t_3d\t_4 \frac{\text{sgn}( 1-\t_4)}{|1-\t_4|^{1/q}\,|\t_3|^{(1+q)/q}} \frac{\text{sgn}(\t_{34})}{|\t_{34}|^{\frac{3q-2}{2q}-h}}\\
k_c^{a,d}
&=\a \int d\t_3d\t_4 \frac{\text{sgn}( 1-\t_4)}{|1-\t_4|^{1/q}\,|\t_3|^{(1+q)/q}} \frac{1}{|\t_{34}|^{\frac{3q-2}{2q}-h}}\,.
\eal
Using the integrals tabulated in appendix~\ref{evss} one finds that the eigenvalues are given by
\bal
k^{s,d}_c(h)  &= \frac{\a\, \p^2   \cos(\p( \frac{q+2}{4q}-\frac{h}{2}))\Gamma(  \frac{q+2}{2q}-h)}{  \sin(\frac{\p}{2q})\Gamma(\frac{1}{q}) \cos(\frac{\p(1+q)}{2q})\Gamma(\frac{ 1+q}{ q})\sin(\p(\frac{3q-2}{4q}-\frac{h}{2}))\Gamma( \frac{3q-2}{2q}-h )}\,,\label{ds}\\
k^{a,d}_c(h)  &= \frac{\a\, \p^2   \sin(\p( \frac{q+2}{4q}-\frac{h}{2}))\Gamma(  \frac{q+2}{2q}-h)}{  \sin(\frac{\p}{2q})\Gamma(\frac{1}{q}) \cos(\frac{\p(1+q)}{2q})\Gamma(\frac{ 1+q}{ q})\cos(\p(\frac{3q-2}{4q}-\frac{h}{2}))\Gamma( \frac{3q-2}{2q}-h )} \,.\label{da}
\eal
These expressions are easily seen to be in complete agreement with eq.~(6.2) of~\cite{Fu:2016vas}.

\subsection{The non-diagonal 4-point kernels}\label{nondiagonalspectrum}

\begin{figure}[t]
  \begin{center}
\includegraphics[width=0.80\linewidth]{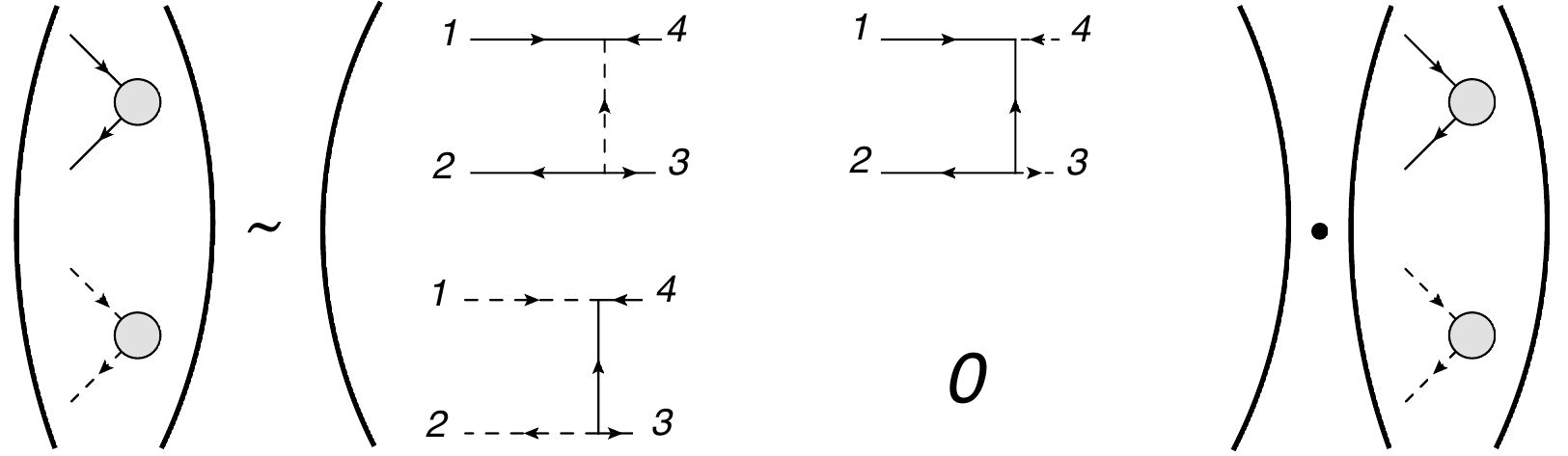}
  \end{center}
\caption{Kernels of the
$\<\psi_i(\t_1)\bar{\psi}_i(\t_2){\psi}_j(\t_3)\bar\psi_j(\t_4)\>$,   $\<\psi_i(\t_1)\bar{\psi}_i(\t_2) {b}_j(\t_3)\bar b_j(\t_4)\>$, $\<b_i(\t_1)\bar{b}_i(\t_2) {\psi}_j(\t_3)\bar \psi_j(\t_4)\>$ and
$\<b_i(\t_1)\bar{b}_i(\t_2) {b}_j(\t_3)\bar b_j(\t_4)\>$ correlation functions. Iterating these kernels to build ladder diagrams amounts to $2 \times 2$ matrix multiplication.}
\label{fig:sk3}
\end{figure}

It is similarly easy to work out kernels corresponding to pairs of bosons or fermions on the same side of the ladder.
These kernels have the property that the two legs on the left are different than the two legs
on the right, so adding any additional such rung to a ladder will change the ``end'' of the latter.
These kernels are therefore assembled into a $2 \times 2$ matrix, as illustrated in figure~\ref{fig:sk3}.

The entries of this kernel matrix $K^{ij}$ are
\bal
K^{11}&=J \frac{(q-1)!}{(q-3)!}G^\psi(\t_{14})G^{\bar\psi}(\t_{23})G^{\bar{b}}(\t_{34})(G^{\psi}(\t_{34}))^{q-3}\,,\\
K^{12}&=J \frac{(q-1)!}{(q-2)!}G^\psi(\t_{14})G^{\bar\psi}(\t_{23})(G^{\psi}(\t_{34}))^{q-2}\,,\\
K^{21}&=J \frac{(q-1)!}{(q-2)!}G^{\bar b}(\t_{14})G^{b}(\t_{23})(G^{\psi}(\t_{34}))^{q-2}\,,
\eal
with $K^{22} = 0$ since there is no connected contribution to
$\<b_i(\t_1)\bar{b}_i(\t_2) {b}_j(\t_3)\bar b_j(\t_4)\>$.
We now repeat the computation of eqs.~\eqref{ees} and~\eqref{eea} to find the
eigenvalues $k^{ij}$ of the three $K^{ij}$.
In the conformal limit we have from eqs.~\eqref{res1} and~\eqref{res2} that
\bal
K_c^{11}&= \frac{(q-2) \a\,\text{sgn}(\t_{14})\,\text{sgn}(\t_{23}) }{|\t_{14}|^{1/q}|\t_{34}|^{(q-3)/q}|\t_{34}|^{(q+1)/q}|\t_{23}|^{1/q}}\,, \nonumber\\
K_c^{12}&=
 \frac{q \a\,\text{sgn}(\t_{14})\,\text{sgn}(\t_{23})\text{sgn}(\t_{34})}{|\t_{14}|^{1/q}|\t_{34}|^{(q-2)/q}|\t_{23}|^{1/q}}\,,\label{matrixkernel}  \\
K_c^{21}&=
 \frac{\frac{\a}{q}\,\text{sgn}(\t_{34})}{|\t_{14}|^{(1+q)/q}|\t_{34}|^{(q-2)/q}|\t_{23}|^{(1+q)/q}}\,.
\nonumber
\eal
As in the diagonal case, the eigenfunctions can be either symmetric or antisymmetric, and we compute them separately.

Here it is important to clarify that we call a 2-component eigenfunction of the kernel matrix ``symmetric'' or ``antisymmetric'' according to the symmetry property of only the first, fermionic component.
The second, bosonic component must have the opposite symmetry, since the off-diagonal kernel entries
$K^{12}_c$ and $K^{21}_c$ are odd under a simultaneous flip
of $\t_1 \leftrightarrow \t_2$ and $\t_3 \leftrightarrow \t_4$.

To find the antisymmetric eigenvalues we therefore consider a 2-component trial
eigenfunction of the form
\bal
\left(\, \frac{\text{sgn}(\t_{34})}{|\t_{34}|^{\frac{1}{q}-h}} \quad \frac{1}{|\t_{34}|^{  \frac{q+1}{ q}-h }} \, \right).\label{asym}
\eal
As in the previous subsection the powers in the denominator are taken in consideration of the
dimensions of the free legs in the
associated kernel diagram, in this case $2 \Delta_\psi$ and $2 \Delta_b$, respectively.
When acting on a vector of this form, we find that the kernel has the non-zero matrix elements
\bal
k_c^{a,11}&=\frac{-(q-2) \a\p^2   \sin(\p(   \frac{1}{2q}-\frac{h}{2} ))\Gamma(  \frac{1}{ q}-h )}{  \sin(\frac{\p}{2q})^2\Gamma(\frac{1}{q})^2  \sin(\p(\frac{2q-1}{2q}-\frac{h}{2}))\Gamma(\frac{2q-1}{q}-h)}\,,\\
k_c^{a,12}&= \frac{-q \a\,\p^2   \sin(\p(\frac{ 1}{ 2q}-\frac{h}{2 } ))\Gamma( \frac{ 1}{ q}-h )}{  \sin(\frac{\p}{2q})^2\Gamma(\frac{1}{ q})^2 \sin(\p( \frac{2q-1}{ 2q}-\frac{h}{2 }))\Gamma( \frac{2q-1}{ q}-h)}\,,  \\
k_c^{a,21}&=\frac{\frac{\a}{q}\p^2   \cos(\p(\frac{q+1}{2q}-\frac{h}{2}))\Gamma(\frac{q+1}{q}-h)}{  \cos(\p\frac{q+1}{2q})^2\Gamma(\frac{q+1}{q})^2 \cos(\p(\frac{q-1}{2q}-\frac{h}{2}))\Gamma(\frac{q-1}{q}-h)}\,.\label{ks21}
\eal
The two eigenvalues of the $2\times 2$ matrix $k_c^{a,ij}$ can be represented as\footnote{The eigenvalues
can also be represented as $\widetilde{k}^{a,\pm}_c = \frac{1}{2} k^{a,11}_c \pm \sqrt{ \frac{1}{4} (k_c^{a,11})^2 + k^{a,12}_c k^{a,21}_c}$. We caution the reader that although the sets $\{ \widetilde{k}^{a,+}_c, \widetilde{k}^{a,-}_c\}$ and $\{ k^{a,+}_c, k^{a,-}_c \}$ are the same for all $h$, it is not true that
$\widetilde{k}^{a,\pm}_c = k^{a,\pm}_c$ for all $h$.}
\begin{equation}
k^{a,\pm}(h)=
\mp \frac{\Gamma(2-\frac{1}{q}) \Gamma(1-\frac{h}{2}-\frac{1}{2q}) \Gamma(\frac{1}{2q} +\frac{h}{2}) \Gamma(\frac{1}{2}-h+\frac{1}{q}\mp \frac{1}{2})}{\Gamma(1+\frac{1}{q})\Gamma(1+\frac{h}{2}-\frac{1}{2q})\Gamma(\frac{1}{2q}-\frac{h}{2}) \Gamma(\frac{3}{2}-h-\frac{1}{q}\mp \frac{1}{2})}\,.
\end{equation}
It can be checked that
\begin{equation}
\label{relation1}
k_c^{a,+}(\textstyle{h-\frac{1}{2}})=k^{a,d}_c (h) \qquad \text{ and } \qquad k_c^{a,-}(\textstyle{h+\frac{1}{2}})=k^{s,d}_c (h)\,,
\end{equation}
in accord with the statement on page 30 of~\cite{Fu:2016vas}.

Similarly, to find the symmetric eigenvalues we act with the kernel on
\begin{equation}
\left( \frac{1}{|\t_{34}|^{\frac{1}{q}-h}} \quad \frac{\text{sgn}(\t_{34})}{|\t_{34}|^{  \frac{q+1}{ q}-h }}\right)\label{sym}
\end{equation}
to find the matrix elements
\bal
k_c^{s,11} &= \frac{(q-2) \a\,\p^2   \cos(\p( \frac{1}{2q}-\frac{h}{2} ))\Gamma( \frac{1}{q}   -h )}{  \sin(  \frac{\p}{2q})^2\Gamma( \frac{1}{q})^2 \cos(\p(\frac{2q-1}{2q}-\frac{h}{2}))\Gamma(\frac{2q-1}{q}-h)}\,,  \\
k_c^{s,12} &=\frac{q \a\,\p^2   \cos(\p( \frac{ 1}{2q}-\frac{h}{2} ))\Gamma(\frac{ 1}{q}-h )}{  \sin( \frac{\p}{2q})^2\Gamma(\frac{1}{q})^2  \cos(\p (\frac{2q-1}{2q}-\frac{h}{2}))\Gamma( \frac{2q-1}{q}-h)}\,,  \\
k_c^{s,21} &=
 \frac{-\frac{\a}{q}\p^2   \sin(\p(\frac{q+1}{2q}-\frac{h}{2} ))\Gamma( \frac{q+1}{q}-h )}{  \cos(\p\frac{1+q}{2q})^2\Gamma(\frac{1+q}{q})^2 \sin(\p(\frac{q-1}{2q}-\frac{h}{2}))\Gamma(\frac{q-1}{q}-h)}\,,\label{ka21}
 \eal
and the corresponding eigenvalues (footnote 1 applies again)
\begin{equation}
k_c^{s,\pm}(h)
= \mp \frac{\Gamma(2-\frac{1}{q}) \Gamma(\frac{1}{2} - \frac{h}{2} - \frac{1}{2q})
\Gamma(\frac{1}{2} + \frac{h}{2} + \frac{1}{2q}) \Gamma(\frac{1}{2}-h+\frac{1}{q} \mp \frac{1}{2})}
{\Gamma(1+\frac{1}{q}) \Gamma(\frac{1}{2}+\frac{h}{2}-\frac{1}{2q}) \Gamma(\frac{1}{2}-\frac{h}{2}+\frac{1}{2q})\Gamma(\frac{3}{2}-h-\frac{1}{q} \mp \frac{1}{2})}\,.
\end{equation}
Notice that
\begin{equation}
\label{relation2}
k_c^{s,+}(\textstyle{h-\frac{1}{2}})=k^{s,d}_c (h) \qquad \text{ and } \qquad k_c^{s,-}(h+\frac{1}{2})=k^{a,d}_c (h)\,,
\end{equation}
again in accord with~\cite{Fu:2016vas}.

\begin{figure}[t!]
    \centering
    \begin{subfigure}[t]{0.45\textwidth}
        \includegraphics[width=\textwidth]{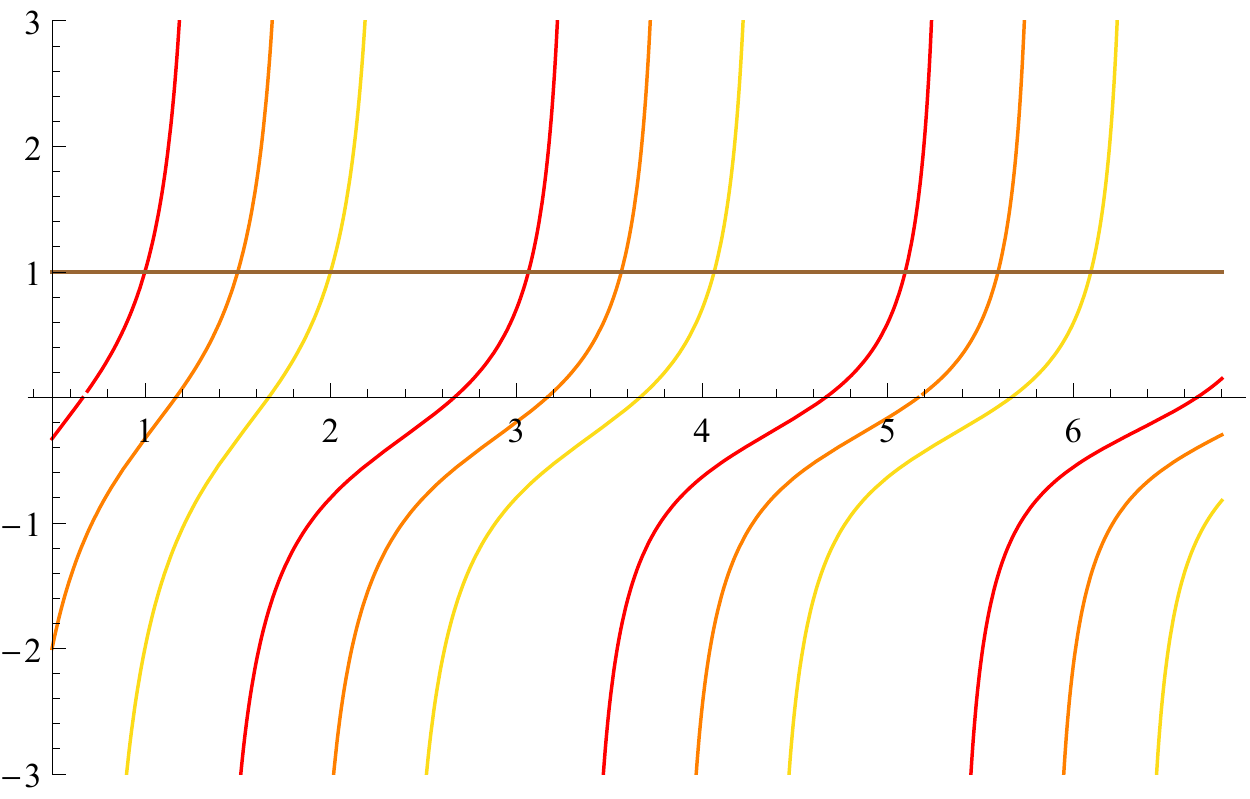}
        \caption{The red, orange and yellow curves represent the eigenvalues $k_c^{a,+}(h)$, $k^{a,d}_c (h)$  and $k_c^{s,-}(h)$ respectively.  }
        \label{fig:ev1a}
    \end{subfigure}
    \quad
    \begin{subfigure}[t]{0.45\textwidth}
        \includegraphics[width=\textwidth]{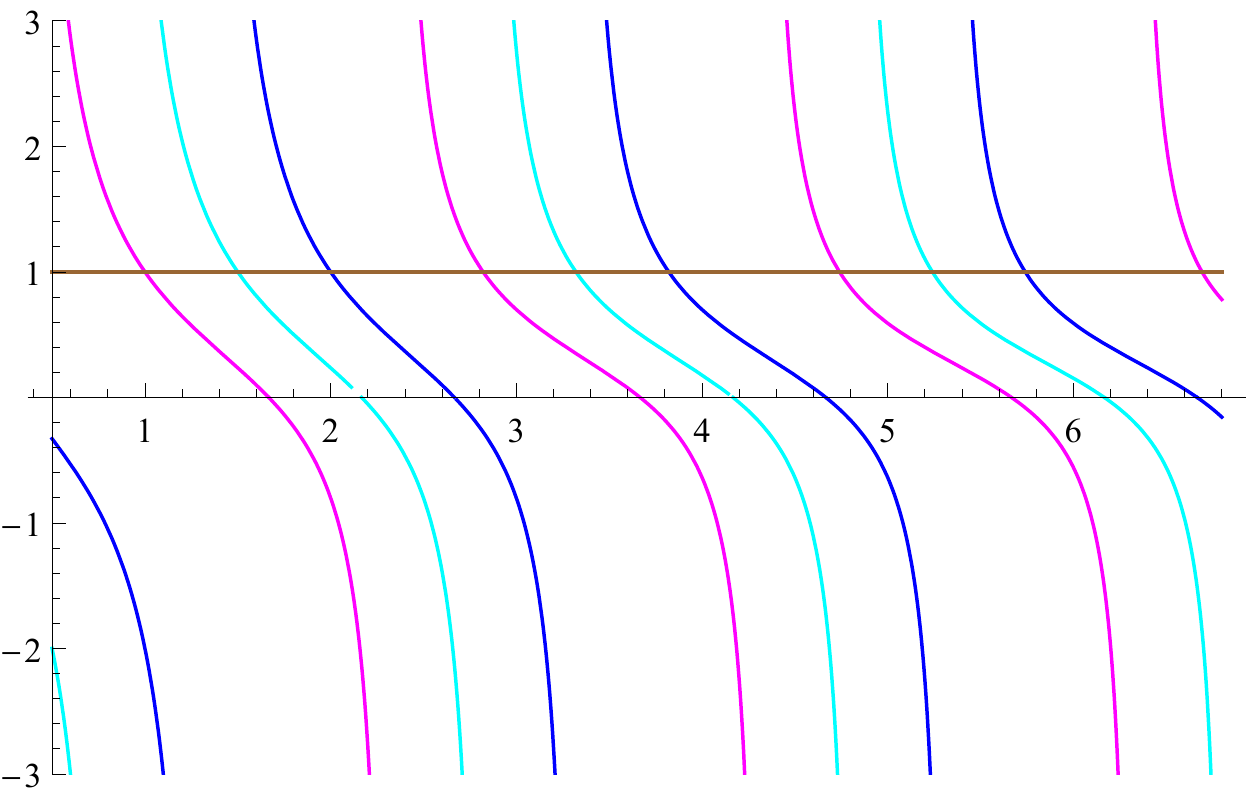}
        \caption{The magenta, cyan and blue curves represent the eigenvalues $k_c^{s,+}(h)$, $k^{s,d}_c (h)$  and $k_c^{a,-}(h)$ respectively.}
        \label{fig:ev1s}
    \end{subfigure}
    \caption{A plot of the eigenvalues for the $q=3$ model. Each figure illustrates one tower of $\cn=2$ supermultiplets. Their intersects with the horizontal $k=1$ line give the dimensions of operators running in the OPE channel. }\label{fig:ev1}
\end{figure}

It should not be a surprise to have found in eq.~\eqref{relation1}
that $k^{a,-}_c$ can be related to $k_c^{s,d}$ instead of to
$k_c^{a,d}$: this is a consequence of the different symmetry properties of the two entries
in~\eqref{asym}.
The relation~\eqref{relation2} between $k^{s,-}_c$ and $k_c^{a,d}$ occurs for the same reason.

For a given kernel $K$, the dimensions of the operators running in the OPE channel of $K$
are the values of $h$ for which the eigenvalue(s) of $K$ satisfy $k(h) = 1$.
From the relation~\eqref{relation1}
between the diagonal and non-diagonal 4-point kernels, it is natural to expect that operators whose dimensions arise from the $k_c^{a,+}(h)$, $k^{a,d}_c (h)$ and $k_c^{s,-}(h)$ eigenvalues assemble into a tower of $\cn=2$ supermultiplets with dimensions $\{h - \frac{1}{2}, h, h + \frac{1}{2}\}$.
For example, for $q=3$ we have
\bal
k_c^{a,+}(h)=1 &\quad \Rightarrow \quad h^{a,+}_m=1 ,\, 3.0659,\, 5.09488,\,7.11311,\,9.12623,\,\ldots \label{hap}\\
k^{a,d}_c (h) =1 &\quad \Rightarrow \quad  h^{a,d}_m= 1.5,\, 3.5659,\, 5.59488,\, 7.61311,\, 9.62623,\,\ldots \label{had}\\
k_c^{s,-}(h)=1 &\quad \Rightarrow \quad   h^{s,-}_m=2,\, 4.0659,\, 6.09488,\, 8.11311,\, 10.1262,\, \ldots \label{hsm}
\eal
in agreement with eq.~(6.3) of~\cite{Fu:2016vas}.
These eigenvalues and dimensions are shown in figure~\ref{fig:ev1a}.

Similarly, the dimensions from the $k_c^{s,+}(h)$, $k^{s,d}_c (h)$ and $k_c^{a,-}(h)$ eigenvalues comprise a second tower of $\cn=2$ supermultiplets. The dimensions of these eigenvalues at $q=3$ are
\bal
k_c^{s,+}(h)=1 &\quad \Rightarrow \quad h^{s,+}_m=1 ,\, 2.82114,\, 4.74091,\, 6.69332,\, 8.66092,\,\ldots \label{hsp}\\
k^{s,d}_c (h) =1 &\quad \Rightarrow \quad h^{s,d}_m= 1.5,\, 3.32114,\, 5.24091,\, 7.19332,\, 9.16092,\,  \ldots \label{hsd}\\
k_c^{a,-}(h)=1 &\quad \Rightarrow \quad   h^{a,-}_m=2,\, 3.82114,\, 5.74091,\,7.69332,\, 9.66092,\, \ldots \label{ham}
\eal
again in agreement with~\cite{Fu:2016vas}.
These eigenvalues and dimensions are shown in figure~\ref{fig:ev1s}.

\subsection{The chaotic behavior}

The chaotic behavior of this model can be studied in a similar way by analyzing the retarded kernels.
In this subsection we carry out this analysis, partly also as a double check of our computations in the previous subsections.

The chaotic behavior is measured by out-of-time-order correlators~\cite{KitaevTalk1,KitaevTalk2}, which can be obtained either from an analytic continuation of the Euclidean 4-point function or directly from analyzing the retarded kernel.  We take the latter approach, following~\cite{KitaevTalk1, KitaevTalk2} and section 3.6.1 of~\cite{Maldacena:2016hyu}. The retarded kernel is defined on a complex time contour with two real time folds on two antipodal points on the thermal circle. It can expressed in terms of the
 retarded and ladder rung propagators
 \bal
 G_R^\psi(t)& =2 \cos(\p\Delta_\psi) b_\psi \Big(\frac{\p }{  \b \sinh(\frac{\p  t }{\b})}\Big)^{2\Delta_\psi}\Theta(t)\,,\\
  G_R^b(t)&=-2i \sin(\p\Delta_b) b_b \Big(\frac{\p }{  \b \sinh(\frac{\p  t }{\b})}\Big)^{2\Delta_b}\Theta(t)\,,\\
G_{lr}^x(t)&= b_x \Big(\frac{\p}{\b\cosh(\frac{\p t}{\b})}\Big)^{2\Delta_x}\,, \qquad x\in\{\psi,b\}\,,
 \eal
 which are obtained from the Euclidean propagators by analytic continuation~\cite{Maldacena:2016hyu,Peng:2017kro}.

The retarded kernel contributions to the 4-point functions shown in figure~\ref{fig:sk3} are
\bal
K_R^{11}&=J \frac{(q-1)!}{(q-3)!}G_R^\psi(t_{14})G_R^{\bar\psi}(t_{23})G_{lr}^{b}(t_{34})(G_{lr}^{\psi}(t_{34}))^{q-3}\,,\\
K_R^{12}&=-J \frac{(q-1)!}{(q-2)!}G_R^\psi(t_{14})G_R^{\bar\psi}(t_{23})(G_{lr}^{\psi}(t_{34}))^{q-2}\,,\\
K_R^{21}&=J \frac{(q-1)!}{(q-2)!}G_R^{\bar b}(t_{14})G_R^{b}(t_{23})(G_{lr}^{\psi}(t_{34}))^{q-2}\,.
\eal
We have included a factor of $i^2$ in each case due to the vertex insertion on the Lorentzian time folds, and we have also included a factor of $-1$ on the kernels $K_R^{11}$, $K^{21}_R$ arising from the operator ordering required by the contour~\cite{Peng:2017kro}.
When acting by matrix multiplication and convolution on an ansatz of the form
\begin{equation}
\left(\frac{e^{-h\frac{\p}{\b} (t_1+t_2) } }{(\frac{\beta}{\pi} \cosh\frac{\p t_{12}}{\b} )^{\frac{1}{q}-h}}\quad
\frac{e^{-h\frac{\p}{\b}  (t_1+t_2)} }{(\frac{\beta}{\pi} \cosh\frac{\p t_{12}}{\b} )^{\frac{1+q}{q}-h}}\right),\label{therazt}
\end{equation}
the nonzero matrix elements of the kernel can be explicitly evaluated as
\bal
k_R^{11}&=4 (q-2)\a    \frac{ \cos^2(\frac{\p}{2q}) \Gamma (\frac{q-1}{q})^2   \Gamma (\frac{1}{q}-h)}{ \Gamma (2-h-\frac{1}{q})}\,,\\
k_R^{12}&=-4 q\a  \frac{\cos^2(\frac{\p}{2q}) \Gamma (\frac{q-1}{q})^2   \Gamma (\frac{1}{q}-h) }{ \Gamma (2-h-\frac{1}{q})}\,,\\
k_R^{21}&=-4 \frac{\a}{q}    \frac{ \cos^2(\frac{\p}{2q}) \Gamma (-\frac{1}{q})^2   \Gamma (1+\frac{1 }{q}-h)  }{ \Gamma (1-h-\frac{1}{q})}\,.
\eal
The two eigenvalues of the matrix $k_R^{ij}$ are
\begin{equation}
k_R^\pm(h) = \mp \frac{\Gamma(2-\frac{1}{q}) \Gamma(\frac{1}{2} - h + \frac{1}{q} \pm \frac{1}{2})}
{\Gamma(1+\frac{1}{q}) \Gamma(\frac{3}{2} - h - \frac{1}{q} \pm \frac{1}{2})}\,.
\end{equation}
The eigenvalue $k_{R}^-$ reaches its resonance value $k_R^-(h) = 1$ at $h=-1$. This indicates chaos: the eigenfunctions~\eqref{therazt} grow exponentially with maximal Lyapunov parameter
 \bal
 \l_L=\frac{2\p}{\b}\,.
 \eal
 There are no other negative values of $h$ that set any of the eigenvalues to one. Therefore there is no ``subleading'' chaotic behavior.

\section{Symmetric and Antisymmetric Conformal Eigenfunctions}

The simplest way of summing ladder diagrams to evaluate a 4-point function is to
first expand the 0-rung ladder diagram in a complete basis of eigenfunctions of the kernels.
It is then straightforward to generate the $L$-rung ladder, and then to sum all ladder diagrams,
since the kernels act (by convolution) diagonally in this basis.  Furthermore one can make
efficient use of conformal symmetry, which effectively reduces all 4-point calculations
to functions of a single cross-ratio $\c$.  In this section, which follows closely
section 3.2 of~\cite{Maldacena:2016hyu}, we determine the complete
set of conformal eigenfunctions of the various ladder kernels.
Particular importance is played by the transformation
$\c \to \frac{\c}{\c-1}$ which is a symmetry of both the fermionic
SYK model studied
in~\cite{KitaevTalk1,KitaevTalk2,Maldacena:2016hyu} and the
$\cn = 1$ supersymmetric model studied
in~\cite{Fu:2016vas,Murugan:2017eto}.
However, in our study of the $\cn = 2$ model we will also require
eigenfunctions that are
antisymmetric under $\c \to \frac{\c}{\c-1}$.

It is straightforward to check that the kernels commute with the conformal Casimir operator, which reads
\bal
\cc=\c^2(1-\c)\pa_\c^2
-\c^2\pa_\c
\eal
in terms of the conformal cross ratio
\begin{equation}
\chi = \frac{\t_{12}\t_{34}}{\t_{13}\t_{24}}\,.
\label{crossratio}
\end{equation}
It admits a set of eigenfunctions $\F_h(\c)$, with eigenvalues $h(h-1)$, that satisfy
\bal
\cc \F_h(\c)=h(h-1)\F_h(\c)\,.
\label{casimir}
\eal
This is an equation of hypergeometric type that for $0<\c<1$ has two linearly independent solutions
\bal
F_1(\c) &= \frac{\c^h {} \Gamma(h)^2}{\Gamma(2h)} \, _2F_1(h,h;2 h;\c)\,,\\
F_2(\c) &= \frac{\c^{1-h} \Gamma(1-h)^2}{\Gamma(2(1-h))} {}\, _2F_1(1-h,1-h;2-2h;\c)
\eal
related by $h \to 1 - h$, where we have chosen an overall normalization for later convenience.
These expressions are useful because they manifest the behaviors $\chi^h$, $\chi^{1-h}$ of the eigenfunctions
near $\chi = 0$, but it is more convenient to work in a different basis of solutions
where the symmetry properties under the transformation $\c \to \frac{\c}{\c-1}$
are manifest.
We will denote the eigenfunctions that are symmetric and antisymmetric
under this transformation by
$\Phi_h^s(\c)$ and $\Phi_h^a(\c)$, respectively.

In the region $\c>1$, the two solutions with definite parity under $\c \to \frac{\c}{\c-1}$ are
\bal
\F^s_{\chi>1}(\c)&=\frac{\Gamma \left(\frac{1}{2}-\frac{h}{2}\right) \Gamma \left(\frac{h}{2}\right)}{\sqrt{\pi }} {}_2F_1\left(\frac{h}{2},\frac{1-h}{2},\frac{1}{2},\frac{(\c-2)^2}{\c^2}\right),\\
\F^a_{\chi>1}(\c)&=-\frac{2 \Gamma \left(1-\frac{h}{2}\right) \Gamma \left(\frac{h}{2}+\frac{1}{2}\right)}{\sqrt{\pi }}\frac{\c-2}{\c }  {}_2F_1\left(\frac{2-h}{2},\frac{h+1}{2},\frac{3}{2},\frac{(\c-2)^2}{\c^2}\right).
\eal
These can be extended to the region $0 < \c < 1$ by matching their behaviors at $\chi \sim 1$ to
appropriate linear combinations of $F_1$ and $F_2$.
Specifically, as discussed in~\cite{Maldacena:2016hyu}, if $f_{\c>1} \sim A + B \log(\chi-1)$
as $\chi \to 1^+$, then the corresponding extension $f_{0<\c<1}$ should approach $A + B \log(1-\chi)$ as
$\chi \to 1^-$.  In this manner we find that
the appropriate expressions in the region $0 < \c < 1$ are
\bal
\F^s_{0<\c<1}(\c)&=A F_1(\c)+BF_2(\c)\,,\\
\F^a_{0<\c<1}(\c)&=B F_1(\c) +AF_2(\c)\,,
\eal
where
\begin{equation}
A=\frac{1}{2} \tan(\pi h) {\textstyle{\cot(\frac{\pi h}{2})}}\,, \qquad
B=-\frac{1}{2} \tan(\pi h) {\textstyle{\tan(\frac{\pi h}{2})}}\,.
\end{equation}

Finally, the eigenfunctions can be extended to the region $\c < 0$ by exploiting the transformation
$\c \to \frac{\c}{\c - 1}$,
\bal
\F^s_{\c<0} = + \F^s_{0<\c<1}(\textstyle{\frac{\c}{\c-1}})\,, \qquad
\F^a_{\c<0} = - \F^a_{0<\c<1}(\textstyle{\frac{\c}{\c-1}})\,.
\eal

Following~\cite{Maldacena:2016hyu}, the range of allowed values of $h$ can be determined by
requiring that the Casimir be hermitian with respect to the inner product
\bal
\<g,f\>=\frac{1}{2}\int_{-\infty}^\infty \frac{d\c}{\c^2} g(\c)^*f(\c)\,.\label{geninner}
\eal
Convergence of this integral requires the eigenfunctions to approach zero at least as fast as $\chi^{1/2}$.
This restricts the set of allowed symmetric and antisymmetric eigenfunctions and eigenvalues to
\bal
\F_h^s(\c) &\quad \text{ with } \quad h = \frac{1}{2} + i s, s > 0, \quad \text{ or } \quad h=2n,  n\in \bz_+,
\label{spectrums}
\\
\F_h^a(\c) &\quad \text{ with } \quad h = \frac{1}{2} + i s, s > 0, \quad \text{ or } \quad h=2n-1,  n\in \bz_+.
\eal

Notice that due to the degeneracy under $h \to 1-h$, which originates from the form of the eigenvalue $h(h-1)$ of the Casimir, we can choose to restrict the
parameters $s$ in the continuous spectra $h=\frac{1}{2}+i s$ to be positive.
The continuous series of eigenfunctions admit useful integral representations
\bal
\F^s_h(\c)&=\frac{1 }{2  }\int_{-\infty }^\infty  \frac{ |\c|^h}{|y|^{ h}|y-1|^{1-h}|y-\c|^h} \, dy\,,
\label{symint}\\
\F^a_h(\c)&=\frac{\text{sgn}(\c) }{2  }\int_{-\infty }^\infty  \frac{ |\c|^h\, \text{sgn}(y)\text{sgn}(y-1)\text{sgn}(y-\c)}{|y|^{ h}|y-1|^{1-h}|y-\c|^h} \, dy\,,  \label{asymint}
\eal
the first of which was pointed out in~\cite{Maldacena:2016hyu}.
These can be explicitly checked by carrying out the integrations in the four different intervals.
The continuous eigenfunctions are orthogonal, with
\bal
\<\F^s_h,\,\F^s_{h'}\>&=\frac{\p \tan(\p h)}{4h-2}2\p\delta(h-h')\,, \nonumber \\
\label{norm1}
\<\F^a_h,\,\F^a_{h'}\>&=\frac{\p \tan(\p h)}{4h-2}2\p\delta(h-h')\,,\\
\<\F^a_h,\,\F^s_{h'}\>&=0=\<\F^s_h,\,\F^a_{h'}\>\,. \nonumber
\eal
The orthogonality between the symmetric and antisymmetric eigenfunctions can be simply understood
as a consequence of the invariance of the measure $d\c/\c^2$ under $\chi \to \frac{\c}{\c-1}$.

The discrete eigenfunctions can be identified as the real part of the Legendre $Q_\n$ functions of the second kind:
\bal
\F^s_h&=2\,{\rm Re}\left(Q_{h-1}({\textstyle{\frac{2-\c}{\c}}})\right),\qquad h=2n,~n\in \bz_+,\\\F^a_h&=2\,{\rm Re}\left(Q_{h-1}({\textstyle{\frac{2-\c}{\c}}})\right),\qquad h=2n-1,~n\in \bz_+,\label{disas}
\eal
which is a straightforward generalization of~\cite{Maldacena:2016hyu}.
The inner product between the discrete eigenfunctions is simply encapsulated in the formula
\bal
\label{norm2}
\<\F_h,\F_{h'}\>=\frac{\p^2\del_{hh'}}{4h-2}\,.
\eal
In particular, the $\del_{hh'}$ implies that the symmetric and antisymmetric eigenfunctions are orthogonal since their $h$ parameters must be even and odd, respectively.

We conclude this section by using the completeness relation to write an explicit formula for the eigenfunction
decomposition of an arbitrary function $f(\chi)$.  Schematically it reads
\begin{equation}
f(\chi) =\, \int\hspace{-6mm}{\sum}_{h,i} \frac{f^i(h)}{\langle \Phi_h^i, \Phi_h^i \rangle}
\Phi_h^i(\chi) \qquad \text{ where } f^i(h) \equiv \langle f, \Phi_h^i\rangle\,,
\end{equation}
where the sum over all
eigenfunctions includes an index
$i \in \{s, a\}$ that accounts for both the symmetric and antisymmetric sectors,
and $\int\hspace{-12.5pt}\mathrel{\raisebox{1.5pt}{$\scriptstyle\sum$}}_{h}$ denotes a sum over the discrete states and an integral over the continuous
series.
Specifically, using eqs.~\eqref{norm1} and~\eqref{norm2}, we have
\begin{equation}
f(\chi) =
\sum_{h=1}^\infty \frac{(4h-2)}{\pi^2} \left\{\begin{matrix}
f^s(h) \Phi_h^s(\chi) \text{ if } h \text{ is even }\\
f^a(h) \Phi_h^a(\chi) \text{ if } h \text{ is odd }
\end{matrix}
\right\}
+
\sum_{i\in \{ s, a \} } \int_0^\infty \frac{ds}{2 \pi} \frac{(4h-2)}{\pi \tan(\pi h)} f^{i}(h) \Phi_h^i(\chi)\,,
\end{equation}
where in the integral it should be understood that $h = \frac{1}{2} + i s$.
Similarly to the case of~\cite{Maldacena:2016hyu}, this formula can be understood as a single
contour integral over a contour in the complex $h$ plane defined as
\begin{equation}
\frac{1}{2 \pi i} \int_\cc dh \equiv \int_{-\infty}^{+\infty} \frac{ds}{2 \pi} + \sum_{n=1}^\infty
{\rm Res}_{h=n}\,.
\end{equation}
In this sense, then, we have finally
\begin{equation}
f(\chi) = \frac{1}{2 \pi i} \int_\cc dh \left( \frac{(2h-1) f^s(h)}{\pi \tan(\frac{\pi h}{2})}
\Phi_h^s(\chi) + \frac{(2h-1) f^a(h)}{\pi \tan(\frac{\pi(h-1)}{2})} \Phi_h^a(\chi) \right).
\label{completeness}
\end{equation}

\section{Evaluating the 4-point Functions}
\label{fourpointsection}

Given the complete set of eigenfunctions of the kernels, we are now ready to compute the full 4-point functions in this model, following closely the similar calculation in the fermionic SYK model~\cite{KitaevTalk1,KitaevTalk2,Maldacena:2016hyu}.
Throughout this analysis we
left out a proper treatment of the (so-called ``enhanced'') contributions
from the lowest $\cn = 2$ supermultiplet, although we check at the very
end of this section that all divergences in the 4-point functions
indeed arise from exchange
of this multiplet.

\subsection{Setup}

There are several independent 4-point functions. We consider first those of the type encountered in section~\ref{nondiagonalspectrum} for which both pairs of $U(N)$ indices are contracted between fields with the same statistics:
\bal
\<\psi_i(\t_1)\bar{\psi}_i(\t_2)\psi_j(\t_3)\bar{\psi}_j(\t_4)\>\,, &\quad
\<b_i(\t_1)\bar{b}_i(\t_2)\psi_j(\t_3)\bar{\psi}_j(\t_4)\>\,,\label{firsttwo}\\
\<\psi_i(\t_1)\bar{\psi}_i(\t_2)b_j(\t_3)\bar{b}_j(\t_4)\>\,, &\quad
\<b_i(\t_1)\bar{b}_i(\t_2)b_j(\t_3)\bar{b}_j(\t_4)\>\,. \label{lasttwo}
\eal
These correlation functions take the form
\bal
&\frac{1}{N^2}\sum_{i,j=1}^N\<\psi_i(\t_1)\bar{\psi}_i(\t_2)\psi_j(\t_3)\bar{\psi}_j(\t_4)\>\\
&\quad\quad\quad\quad\quad=G^\psi(\t_{12})G^\psi(\t_{34})+\frac{1}{N} \cf^{\psi\bar{\psi}\psi\bar{\psi}}(\t_1,\t_2,\t_3,\t_4)+\co(\frac{1}{N^2})\label{ftime}\\
&\quad\quad\quad\quad\quad=G^\psi(\t_{12})G^\psi(\t_{34})\left[1+\frac{1}{N} \cf^{\psi\bar{\psi}\psi\bar{\psi}}(\c)+\co(\frac{1}{N^2})\right]\label{fcro},
\eal
and similarly for the others.
We would like to compute the $\cf^{x\bar{x}y\bar{y}}$'s. Notice that we have adopted two conventions for $\cf^{x\bar{x}y\bar{y}}$ and it is meant to
  be understood that when we write the argument as the cross-ratio $\c$ defined in eq.~\eqref{crossratio},
we mean the function
in~\eqref{fcro}, and when we write the arguments as $\t_1 ,  \t_2 ,  \t_3 ,  \t_4$,
  we mean the function
in~\eqref{ftime}.
The zero-rung (tree-level) contributions to these 4-point functions, which we indicate
with a subscript ``0'', are
\bal
\cf^{\psi\bar{\psi}\psi\bar{\psi}}_0(\t_1,\t_2,\t_3,\t_4) &= G^\psi(\t_{14}) G^\psi(\t_{23})\,,\\
\cf^{b\bar{b}b\bar{b}}_0(\t_1,\t_2,\t_3,\t_4) &= G^b(\t_{14}) G^b(\t_{23})\,,\\
\cf^{\psi\bar{\psi}b\bar{b}}_0(\t_1,\t_2,\t_3,\t_4) &= \cf^{b\bar{b}\psi\bar{\psi}}_0(\t_1,\t_2,\t_3,\t_4) = 0\,.
\eal
In terms of the cross ratio, the non-zero ones read
\bal
\cf_0^{\psi\bar{\psi}\psi\bar{\psi}}(\c)
= \text{sgn}\left(\frac{\c}{1-\c}\right)\left|\frac{\c}{1-\c}\right|^{\frac{1}{q}},\qquad
\cf_0^{b\bar{b}b\bar{b}}(\c)
= \left|\frac{\c}{1-\c}\right|^{\frac{1+q}{q}}.\label{F0psi}
\eal

Next we decompose the 0-rung correlators~\eqref{F0psi} into the conformal eigenfunctions constructed
in the previous section.
To this end we first compute the inner products of the 0-rung correlators~\eqref{F0psi}
with the symmetric and antisymmetric eigenfunctions, using their integral
representations~\eqref{symint} and~\eqref{asymint}.
These integrals can be evaluated straightforwardly using formula (3.11)
of~\cite{Peng:2017kro} (see also the appendix for more details on these types of integrals),
and we find
\bal
\<\cf^{\psi\bar{\psi}\psi\bar{\psi}}_0,\F^s_h\>
&=-\frac{\pi  \cos ^2\left(\frac{\pi }{2 q}\right) \Gamma \left(\frac{q-1}{q}\right)^2}{\left(\sin (\pi  h)+\sin \left(\frac{\pi }{q}\right)\right) \Gamma \left(-h-\frac{1}{q}+2\right) \Gamma \left(h-\frac{1}{q}+1\right)}
\equiv k^{\psi,s}_0(h)\,, \\
\<\cf^{\psi\bar{\psi}\psi\bar{\psi}}_0,\F^a_h\>
&=\frac{\pi  \cos ^2\left(\frac{\pi }{2 q}\right) \Gamma \left(\frac{q-1}{q}\right)^2}{ \left(\sin \left(\frac{\pi }{q}\right)-\sin (\pi  h)\right) \Gamma \left(-h-\frac{1}{q}+2\right) \Gamma \left(h-\frac{1}{q}+1\right)}
\equiv k^{\psi,a}_0(h)\,,
\eal
\bal
\<\cf^{b\bar{b}b\bar{b}}_0,\F^s_h\>
&=\frac{\pi  \cos ^2\left(\frac{\pi }{2 q}\right) \Gamma \left(-\frac{1}{q}\right)^2}{\left(\sin (\pi  h)+\sin \left(\frac{\pi }{q}\right)\right) \Gamma \left(-h-\frac{1}{q}+1\right) \Gamma \left(h-\frac{1}{q}\right)}
\equiv k^{b,s}_0(h)\,,\\
\<\cf^{b\bar{b}b\bar{b}}_0,\F^a_h\>
&= \frac{\pi  \cos ^2\left(\frac{\pi }{2 q}\right) \Gamma \left(-\frac{1}{q}\right)^2}{\left(\sin (\pi  h)-\sin \left(\frac{\pi }{q}\right)\right) \Gamma \left(-h-\frac{1}{q}+1\right) \Gamma \left(h-\frac{1}{q}\right)}
\equiv k^{b,a}_0(h)\,.
\eal

The two 4-point functions in eq.~\eqref{firsttwo} are closed under iterating the kernel
shown in figure~\ref{fig:sk3}, as are the two shown in eq.~\eqref{lasttwo}.
We consider the two pairs in turn, since the calculations are essentially identical.

\subsection{The $\psi \bar{\psi} \psi \bar{\psi}$ and $b \bar{b} \psi \bar{\psi}$ 4-point functions}\label{psi4}

The eigenfunction expansion of $\cf^{\psi\bar{\psi}\psi\bar{\psi}}_0(\chi)$
is given by
plugging $k_0^{\psi,s}(h)$ and $k_0^{\psi,a}(h)$ into the completeness relation~\eqref{completeness}.
We can write this as
\begin{equation}
\cf^{\psi\bar{\psi}\psi\bar{\psi}}_0(\chi) = \frac{1}{2 \pi i} \int_\cc dh \left(f_0^{s}(h)+f_0^{a}(h)\right)\\
\end{equation}
where
\begin{equation}
f_0^{s}(h)=  \frac{(2h-1)k^{\psi,s}_0(h)}{\p\tan(\frac{\p h}{2})} \F^s_h(\c)\,, \qquad f_0^{a}(h) = \frac{(2h-1)k^{\psi,a}_0(h)}{\p\tan(\frac{\pi(h-1)}{2})}  \F^a_h(\c)\,.
\end{equation}
Although $\cf^{b\bar{b}\psi\bar{\psi}}_0(\chi) = 0$, let us still write it as
\begin{equation}
\cf^{b\bar{b}\psi\bar{\psi}}_0(\c)=\frac{1}{2\p i}\int_\cc dh \left(m^s_0(h)+m^a_0(h)\right), \quad m^{s,a}_0(h)=0\,,
\end{equation}
so that it serves as a starting point of the repeated action by the kernel matrix.

Starting from the zero-rung ladders characterized by
$f_0^{s,a}(h)$ and $m_0^{s,a}(h)$, we generate the $n$-rung ladder contribution
to the 4-point functions by convolving $n$ times with the kernel matrix~\eqref{matrixkernel}.
The full 4-point functions are then determined by the geometric series
\begin{equation}
f^{s,a}(h) = \sum_{n=0}^\infty f_n^{s,a}(h)\,, \qquad
m^{s,a}(h) = \sum_{n=0}^\infty m_n^{s,a}(h)\,.
\label{laddersum}
\end{equation}
From our previous analysis and the integral relations
\bal
\frac{\text{sgn}(\t_{12})\text{sgn}(\t_{34})}{|\t_{12}|^{2\Delta}|\t_{34}|^{2\Delta}}\F^s_h&=
+\frac{1}{2}\int d\t_0 \frac{ \text{sgn}(\t_{12})}{|\t_{10}|^{h } |\t_{20}|^{h } |\t_{12}|^{2\Delta-h}} \frac{\text{sgn}(\t_{34})}{|\t_{30}|^{1-h} |\t_{40}|^{1-h} |\t_{34}|^{2\Delta -1+h}}\,,\\
\frac{\text{sgn}(\t_{12})\text{sgn}(\t_{34})}{|\t_{12}|^{2\Delta}|\t_{34}|^{2\Delta}}\F^a_h&=
-\frac{1}{2}\int d\t_0 \frac{ \text{sgn}(\t_{01}) \text{sgn}(\t_{02})}{|\t_{10}|^{h } |\t_{20}|^{h } |\t_{12}|^{2\Delta-h}} \frac{\text{sgn}(\t_{03})\text{sgn}(\t_{04})}{|\t_{30}|^{1-h} |\t_{40}|^{1-h} |\t_{34}|^{2\Delta -1+h}}\,,
\eal
we conclude that $(f^s_n , m^s_n)$ close under the action of the kernels $K_c^{11}, K_c^{12}, K_c^{21}$ with eigenvalues $k_c^{a,11},k_c^{a,12},k_c^{a,21}$ respectively; while $(f^a_n, m^a_n)$ is another basis on which the kernels $K^{11}, K^{12}, K^{21}$ have eigenvalues  $k_c^{s,11},k_c^{s,12},k_c^{s,21}$ respectively.

Let us consider the symmetric sector first; the action of the kernel matrix on an $n$-rung ladder is
\bal
 \begin{pmatrix}
    f^s_{n+1}\\
    m^s_{n+1}
    \end{pmatrix}
    =
      \begin{pmatrix}
    k_c^{a,11}&k_c^{a,12}\\
    k_c^{a,21}& 0
    \end{pmatrix}
    \begin{pmatrix}
    f^s_n\\
    m^s_n
    \end{pmatrix}.
\eal
We can diagonalize the matrix as
\bal
 \begin{pmatrix}
    f^s_{n+1}\\
    m^s_{n+1}
    \end{pmatrix}
 = U^{a\dagger}\begin{pmatrix}
    \widetilde{k}_c^{a,-}&(k_c^{a,21}-k_c^{a,12})\text{sgn}(k_c^{a,21})\\
    0 & \widetilde{k}_c^{a,+}
    \end{pmatrix}
 U^a \begin{pmatrix}
    f^s_n\\
    m^s_n
    \end{pmatrix} \,,
\eal
where the $\widetilde{k}$'s are defined in footnote 1, and the transformation matrix
is~$U^a=U(k_c^{a,11},k_c^{a,12},k_c^{a,21})$ with
\bal
U(x,y,z)=\frac{1}{\sqrt{-x \sqrt{x^2+4 y z}+x^2+2 y z+2 z^2}}\left(
\begin{array}{cc}
 -\frac{\sqrt{x^2+4 y z}-x}{\sqrt{2}\,\text{sgn}(z)} & \sqrt{2} |z| \\
 \sqrt{2} z & \frac{\sqrt{x^2+4 y z}-x}{\sqrt{2}} \\
\end{array}
\right).
\eal
The contribution from the $n$-rung ladder diagram can then be solved straightforwardly:
\bal
 \begin{pmatrix}
    f^s_n\\
    m^s_n
    \end{pmatrix}=U^{a,\dagger}\begin{pmatrix}
    (\widetilde{k}_c^{a,-})^n&\quad (k_c^{a,21}-k_c^{a,12})\frac{(\widetilde{k}_c^{a,-})^n-(\widetilde{k}_c^{a,+})^n}{\widetilde{k}_c^{a,-}-\widetilde{k}_c^{a,+}}\, \text{sgn}(k_c^{a,21})\\
    0 & \quad (\widetilde{k}_c^{a,+})^n
    \end{pmatrix}U^{a}
 \begin{pmatrix}
    f^s_0\\
    m^s_0
    \end{pmatrix}.
\eal
Summing over all such diagrams as in eq.~\eqref{laddersum} then gives
\bal
 \begin{pmatrix}
    f^s \\
    m^s
    \end{pmatrix}
    =
  U^{a,\dagger}  \begin{pmatrix}
   \frac{1}{1- \tilde{k}_c^{a,-}}&\quad \frac{k_c^{a,21}-k_c^{a,12}}{(1- \tilde{k}_c^{a,-})(1- \tilde{k}_c^{a,+})}\, \text{sgn}(k_c^{a,21})\\
    0 & \quad \frac{1}{1-\tilde{k}_c^{a,+}}
    \end{pmatrix}
U^{a } \begin{pmatrix}
    f^s_0\\
    m^s_0
    \end{pmatrix}.
\eal
Multiplying through by the $U$ matrices leads to,
\bal
 \begin{pmatrix}
    f^s\\
    m^s
    \end{pmatrix}
    =\frac{1}{(1- \tilde{k}_c^{a,-})(1- \tilde{k}_c^{a,+})}
   \begin{pmatrix}
   1&\quad k_c^{a,12} \\
    k_c^{a,21} & \quad 1-k_c^{a,11}
    \end{pmatrix}
 \begin{pmatrix}
    f^s_0\\
    m^s_0
    \end{pmatrix}.
\label{throughhere}
\eal
Finally, since $m^s_0 = 0$, we have simply
\bal
f^s(h)=\frac{f_0^s(h)  }{(1- \widetilde{k}_c^{a,-})(1- \widetilde{k}_c^{a,+})}\,,\qquad
m^s(h)=\frac{k_c^{a,21}f_0^s(h) }{(1- \widetilde{k}_c^{a,-})(1- \widetilde{k}_c^{a,+})}\,.\label{fm}
\eal
The calculation in the antisymmetric sector
proceeds in the same way, leading to
exactly the same result but with the ``$s$'' and ``$a$'' superscripts exchanged.

The full 4-point functions are then given by
\begin{equation}
\cf^{\psi\bar{\psi}\psi\bar{\psi}}(\c)=\frac{1}{2\p i}\int_\cc dh \left(f^{s}(h)+f^{a}(h)\right),\quad
\cf^{b\bar{b}\psi\bar{\psi}}(\c)=\frac{1}{2\p i}\int_\cc dh \left(m^s (h)+m^a (h)\right).
\end{equation}
As in the pure fermionic case (see~\cite{Maldacena:2016hyu} for details), we can move the contour in the positive real direction to pick up only the contributions from the poles at $\widetilde{k}_c^{i,+}=1$---the factors $1-\widetilde{k}_c^{i,-}$ in the denominator never vanish since $\widetilde{k}_c^{i,-}$ is always negative, and the poles of the $k_c^{i,21}$ factors in the numerator are all cancelled by poles in the denominator.
In the $\c>1$ region this contour deformation is straightforward and leads to
\bal
\cf^{\psi\bar{\psi}\psi\bar{\psi}}(\c)&=-\sum_{i \in \{s,a\}}\sum_m
 \underset{h=\widetilde{h}_m^{i,+}}{\mathrm{Res}}\left(\frac{f_0^{\overline{i}}(h)  }{(1- \widetilde{k}_c^{i,-})(1- \widetilde{k}_c^{i,+})}\right), \quad \c > 1\,,\label{form1}\\
 \cf^{b\bar{b}\psi\bar{\psi}}(\c)&=-\sum_{i \in \{s,a\}}\sum_m
 \underset{h=\widetilde{h}_m^{i,+}}{\mathrm{Res}}\left( \frac{k_c^{i,21}f_0^{\overline{i}}(h)  }{(1- \widetilde{k}_c^{i,-})(1- \widetilde{k}_c^{i,+})}\right), \quad \c > 1\,,
\label{form2}
 \eal
where the sums run over the roots of $1 - \widetilde{k}^{i,+}_c(h) = 0$, enumerated
here as $\widetilde{h}_m^{i,+}$, and $\overline{i}$ means the complement of $i$ in the set $\{s,a\}$.
In the $\c<1$ region, we meet a similar problem of negative entries of the hypergeometric function as that encountered in~\cite{Maldacena:2016hyu}. It is straightforward to generalize their treatment to our case; the net effect  is to replace the $\F^{a,s}_h$ by $\frac{ \c^h \Gamma (h)^2 }{\Gamma (2 h)}{}\, _2F_1(h,h;2 h;\c)$.
Therefore,
the formulas~\eqref{form1} and~\eqref{form2}
can be extended to the region $0 < \c < 1$ by replacing the $f_0^i(h)$'s with
\bal
\widetilde{f}_0^{s}(h) &=  \frac{(2h-1)k^{\psi,s}_0(h)}{\p\tan(\frac{\p h}{2})} \frac{ \c^h \Gamma (h)^2 }{\Gamma (2 h)}{}\, _2F_1(h,h;2 h;\c)\,,\\
\widetilde{f}_0^{a}(h) &= \frac{(2h-1)k^{\psi,a}_0(h)}{\p\tan(\frac{\pi(h-1)}{2})}  \frac{ \c^h \Gamma (h)^2 }{\Gamma (2 h)}{}\, _2F_1(h,h;2 h;\c)\,.
\eal

\subsection{The $\psi\bar{\psi}b\bar{b}$ and $b \bar{b} b \bar{b}$ 4-point functions}

The other two 4-point functions can be computed similarly.
We begin with the eigenfunction decompositions
of the 0-rung ladders
\bal
\cf^{b\bar{b}b\bar{b}}_0(\c)&=\frac{1}{2\p i}\int_\cc dh \left(b_0^{s}(h)+b_0^{a}(h)\right),\\
\cf^{\psi\bar{\psi}b\bar{b}}_0(\c)&=\frac{1}{2\p i}\int_\cc dh \left(p^s_0(h)+p^a_0(h)\right),
\eal
with
\begin{equation}
b_0^{s}(h)=  \frac{(2h-1)k^{b,s}_0}{\p\tan(\frac{\p h}{2})} \F^s_h(\c)\,,\quad
  b_0^{a}(h) =\frac{(2h-1)k^{b,a}_0}{\p\tan(\frac{\p (h-1)}{2})}  \F^a_h(\c)\,,\quad
p_0^{s,a}(h) = 0\,.
\label{plugin}
\end{equation}
The calculation proceeds the same as in the previous subsection,
except with $(f, m) \mapsto (p, b)$, all the way through eq.~\eqref{throughhere}.
At that step we plug in eq.~\eqref{plugin} which leads to
\bal
p^s(h)=\frac{k_c^{a,12} b^s_0(h)}{(1- \widetilde{k}_c^{a,-})(1- \widetilde{k}_c^{a,+})}\,,\qquad
b^s(h)=\frac{ (1-k_c^{a,11}) b^s_0(h)}{(1- \widetilde{k}_c^{a,-})(1- \widetilde{k}_c^{a,+})}\,.\label{pb}
\eal
again together with the same equation but with  the ``$s$'' and ``$a$'' superscripts exchanged.

With these results, the full 4-point functions are then
\bal
&\cf^{\psi\bar{\psi}b\bar{b}} (\c) =
 -\sum_{i \in \{s,a\}} \sum_m
 \underset{h=\widetilde{h}_m^{i,+}}{\mathrm{Res}}\left( \frac{ k_c^{i,12} b^{\overline{i}}_0(h)}{(1- \widetilde{k}_c^{i,-})(1- \widetilde{k}_c^{i,+})}\right), \quad \c>1\,,\\
&\cf^{b\bar{b}b\bar{b}} (\c) =-\sum_{i\in \{s,a\}} \sum_m
 \underset{h=\widetilde{h}_m^{i,+}}{\mathrm{Res}}\left( \frac{ (1-k_c^{i,11}) b^{\overline{i}}_0(h)}{(1- \widetilde{k}_c^{i,-})(1- \widetilde{k}_c^{i,+})}\right), \quad \c>1\,.
\eal
As we saw in the previous subsection, these formulas can be extended to the region $0<\c<1$ by replacing
the $b^i_0(h)$'s with
\bal
\widetilde{b}_0^{s}(h) &=  \frac{(2h-1)k^{b,s}_0(h)}{\p\tan(\frac{\p h}{2})} \frac{ \c^h \Gamma (h)^2 }{\Gamma (2 h)}{}\, _2F_1(h,h;2 h;\c)\,,\\
\widetilde{b}_0^{a}(h) &= \frac{(2h-1)k^{b,a}_0(h)}{\p\tan(\frac{\p (h-1)}{2})}  \frac{ \c^h \Gamma (h)^2 }{\Gamma (2 h)}{}\, _2F_1(h,h;2 h;\c)\,.
\eal

\subsection{The $\psi b \bar{\psi} \bar{b}$ 4-point function}

Finally we turn to the $\<\psi_i(\t_1)b_i(\t_2)\bar\psi_j(\t_3)\bar{b}_j(\t_4)\>$ 4-point function,
which is more subtle.  It has no disconnected contributions, taking the form
\bal
\frac{1}{N^2}\sum_{i,j=1}^N\<\psi_i(\t_1)b_i(\t_2)\bar{\psi}_j(\t_3)\bar{b}_j(\t_4)\>&=\frac{1}{N} \cf^{\psi b\bar{\psi} \bar{b}}(\t_1,\t_2,\t_3,\t_4)+\co(\frac{1}{N^2})\,,
\eal
with the 0-rung correlator being simply
\begin{equation}
\label{bad}
\cf_0^{\psi b\bar{\psi} \bar{b}}(\t_1,\t_2,\t_3,\t_4) = G^\psi(\t_{13})G^b(\t_{24})=\frac{1}{q}\Big(\frac{\tan(\frac{\p}{2q})}{2\p J}\Big)^{2/q}\frac{\text{sgn}(\t_{13})}{|\t_{13}|^{\frac{1}{q}}}\frac{1}{|\t_{24}|^{\frac{1+q}{q}}}\,.
\end{equation}
We would like to continue working with the conformal eigenfunctions from
section~3.  In order to do this we would have to factor out some appropriate
overall $\tau$ dependence in order to render
eq.~\eqref{bad}
a function of the cross-ratio $\chi$, as for example between eqs.~\eqref{ftime} and~\eqref{fcro}.
We can't, however, divide by the obvious candidate $G^\psi(\t_{13})G^b(\t_{24})$ as this
would give us $\cf_0(\chi) = 1$, and the function ``1'' is not in the allowed spectrum;
it would correspond to the discrete state $h=0$ in eq.~\eqref{spectrums}, which is absent
because it is not normalizable with respect to~\eqref{geninner}.

Let us instead notice that the tree-level 4-point function can be expressed as
\bal
\cf_0^{\psi b\bar\psi\bar b}(\t_1,\t_2,\t_3,\t_4)=\pa_{\t_4} \cg_0(\t_1,\t_2,\t_3,\t_4)
\eal
in terms of the auxiliary quantity
\bal
\label{cgdef}
\cg_0(\t_1,\t_2,\t_3,\t_4)&= \Big(\frac{\tan(\frac{\p}{2q})}{2\p J}\Big)^{2/q}\frac{\text{sgn}(\t_{13})\text{sgn}(\t_{24})}{|\t_{13}|^{\frac{1}{q}}|\t_{24}|^{\frac{1}{q}}}\,.
\eal
This derivative might introduce a spurious $\delta(\t_{24})$ contact term, but since we are not
interested in any such terms we can in practice just commit ourselves to neglecting all possible contact terms at the
very end of any calculation.
It is evident from the powers of the denominator factors in eq.~\eqref{cgdef} that we should
think of $\cg$ roughly as a four-fermion correlator; this will tell us, in particular, the conformal
of the eigenfunctions we should use when diagonalizing the relevant kernel.

Now consider the action of some kernel $K$ on $\cg_0$, defined by
\bal
\cg_1(\t_1,\t_2,\t_3,\t_4)=\int d\t_5d\t_6 K(\t_1,\t_2;\t_5,\t_6) \cg_0(\t_5,\t_6,\t_3,\t_4)\,.
\eal
It is evident that taking a $\t_4$ derivative commutes with the action of the ladder kernel, and
this property clearly extends to arbitrary order as we iterate the kernel.
Therefore, the full 4-point function $\cf^{\psi b\bar\psi\bar b}$
can be obtained by first computing the sum of all ladder contributions to $\cg$ using the
kernel~\eqref{conformalKd} and then
taking $\partial_{\t_4}$.
To compute the latter we begin by
constructing an appropriate function of the cross-ratio $\chi$ by factoring out the same
prefactor as in the four-fermion function~\eqref{fcro}, defining
\bal
\cg_0(\c)&\equiv\frac{\cg_0(\t_1,\t_2,\t_3,\t_4)}{G^\psi(\t_{12})G^\psi(\t_{34})}
=\text{sgn}(\c) |\c|^{\frac{1}{q}}\,.
\eal

Now we can decompose $\cg_0(\chi)$ into the conformal eigenfunctions by plugging
its matrix elements
\bal
\<\cg_0,\F^s_h\>
&=\frac{\pi \cos ^2\left(\frac{\pi }{2 q}\right) \Gamma \left(\frac{q-1}{q}\right)^2}{ \left(\sin (\pi  h)+\sin \left(\frac{\pi }{q}\right)\right) \Gamma \left(-h-\frac{1}{q}+2\right) \Gamma \left(h-\frac{1}{q}+1\right)}
=k^{s}_0(h)\,,\\
\<\cg_0,\F^a_h\>
&=\frac{\pi  \cos ^2\left(\frac{\pi }{2 q}\right) \Gamma \left(\frac{q-1}{q}\right)^2}{\left(\sin \left(\frac{\pi }{q}\right)-\sin (\pi  h)\right) \Gamma \left(-h-\frac{1}{q}+2\right) \Gamma \left(h-\frac{1}{q}+1\right)}
=k^{a}_0(h)
\eal
into the completness relation~\eqref{completeness}.

Since the relevant kernel $K_c^d$ is diagonal, much of the complication encountered in the previous
two subsections is avoided.
The sum over ladder diagrams just inserts a geometric series factor $1/(1-k^{i}(h))$
into the conformal eigenfunction
decomposition, where $k^{i}(h)$  are the eigenvalues of the kernel.
However, we should not use the formulas~\eqref{ds} and~\eqref{da} since those are the eigenvalues
of $K_c^d$ when acting on eigenfunctions of the form~\eqref{ees} and~\eqref{eea}.
Since we are iterating the action of $K_c^d$ on $\cg$, which should be treated like a four-fermion
correlator, we must work out the eigenvalues of $K_c^d$ when acting on
eigenfunctions of the form~\eqref{ees} and~\eqref{eea} with $2 \Delta_\psi$ replacing
$\Delta_\psi + \Delta_b$.
This is readily accomplished by substituting $h \to h + \frac{1}{2}$ in eqs.~(\ref{ees}) and~(\ref{eea}), and hence also into
eqs.~(\ref{ds}) and~(\ref{da}).  The resulting
symmetric and antisymmetric eigenvalues
can be expressed in terms of the matrix elements of $\cg_0$ as
\begin{align}
k_c^{o,s}(h) &= k_c^{s,d}(h + \frac{1}{2}) =
4 \alpha q (h - 1 + \frac{1}{q}) k_0^a(h)\,, \cr
k_c^{o,a}(h) &= k_c^{a,d}(h + \frac{1}{2}) =
4 \alpha q (h - 1 + \frac{1}{q}) k_0^s(h)\,.
\label{kodef}
\end{align}

Before proceeding, however, we should make here one
comment about the operator spectrum in this channel. The dimensions of the  operators running in this channel are encoded in the divergences of the 4-point function, which occur at the values of $h$ such that
\bal
k^{o,i}_c(h)=1\,, \qquad i \in \{ s, a\}\,.\label{req}
\eal
However, this $h$ itself does not directly correspond to the operators in the original 4-point function $\cf^{\psi b\psi b}$.
To see this, let us recall that as noted by~\cite{KitaevTalk2, Maldacena:2016hyu, Gross:2016kjj},
the eigenfunctions are naturally dressed with additional factors that allow them to be expressed as 3-point functions whose form is dictated by conformal
symmetry.  We have used this implicitly in previous sections but here we must be more explicit, writing the eigenfunction for example as eq.~(3.69) of~\cite{Maldacena:2016hyu}:
\bal
\frac{ \text{sgn}(\t_{12})}{|\t_{10}|^{h_o-1/2 } |\t_{20}|^{h_o +1/2} |\t_{12}|^{\Delta_\psi+\Delta_b-h_o}}\,,
\label{wrong}
\eal
where $h^o$ is the dimension of the 3rd operator propagating in the channel.

But, as we have already mentioned,
in the present computation, the appropriate eigenfunctions are not eq.~\eqref{wrong} but
rather
\bal
\frac{ \text{sgn}(\t_{12})}{|\t_{10}|^{h  } |\t_{20}|^{h } |\t_{12}|^{2\Delta_\psi-h}}\,.
\eal
Comparing the two expressions, in particular the power of the last term\footnote{The seeming mismatch of the power of $|\t_{20}|$ does not matter since we can always move $\t_0$ to infinity using conformal invariance, without affecting the result of the calculation. Another way to understand this is to notice that the power of $\t_2$ will be corrected once we take the $\t_4$ derivative; this will produce an extra power of $1/\t_{24}$ in the end.}, which is
the one that
enters directly into the integrals that compute the eigenvalues (c.f. eqs~\eqref{ees} and~\eqref{eea}),
we see that the real dimension $h^o$ of the operator running in the original 4-point function $\cf^{\psi b \bar \psi \bar b}$ is related to $h$ by
\bal
h=h_o-\frac{1}{2} \,.\label{shift}
\eal
Indeed, as a check of this relation, we see that the solution to~\eqref{req}, together with the shift~\eqref{shift}, does give the correct dimensions~\eqref{ds}, \eqref{da} (as manifested already in~\eqref{kodef}).

With this important comment out of the way, we are ready to write
the sum over all ladder diagrams
\begin{equation}
        \cg(\c)=\sum_{i \in \{ s, a \}} \frac{1}{2\p i}\int_\cc dh
\frac{g^{\bar{i}}(h)}{1-k_c^{o,i}(h)} \Phi_h^{\bar{i}}\,,\label{gf}
\end{equation}
where
\begin{equation}
 g^s(h) = \frac{(2h-1)k^{s}_0}{\p\tan(\frac{\p h}{2})}\,,\qquad
g^a(h) = \frac{(2h-1)k^{a}_0}{\p\tan(\frac{\p(h-1)}{2})}\,.
\end{equation}
Notice however that $h=1$ is a solution to both $k^{o,a}_c(h)=1$  and $k^{o,s}_c(h)=1$, therefore in the discrete sum of the odd series $\F^a_h$, the $h=1$ term diverges. This corresponds to an enhancement due to a mode of dimension $h_o=\frac{3}{2}$ that is is similar to the $h=2$ enhancement appearing in the fermionic SYK model~\cite{Maldacena:2016hyu,Fu:2016vas}. In the following computation,  we have excluded such enhanced contributions.

We can then push the contour to the right, picking up contributions from poles of the factor $\frac{k^{\bar{i}}_0}{1-k^{o,i}_c(h)}$.
Due to the proportionality~\eqref{kodef} between $k^{\bar{i}}_0$ and $k^{o,i}_c$, the poles of the numerator $k^{\bar{i}}_0$ are cancelled by poles of $k^{o,i}_c$ in the denominator and thus the only contribution comes from solutions to $k^{o,i}_c(h)=1$.  Furthermore, in light of eq.~\eqref{kodef}, these poles
correspond to the set $h_m^{\overline{i},d}-\frac{1}{2}$, where $h_m^{\overline{i},d}$ are the solutions to
$k_c^{\bar{i},d}(h) = 1$, i.e. the dimensions of states propagating in the OPE channel of the diagonal kernel.
For the $\c>1$ region, this contour manipulation is straightforward and leads to
\begin{equation}
\cg(\c)
=-\sum_{i \in \{s,a\}} \sum_m
\underset{h=h_m^{\bar{i},d}-\frac{1}{2}}{\mathrm{Res}}
\left( \frac{g^{\bar{i}}(h)}{1-k^{o,i}_c(h)}\F^{\bar{i}}_h\right), \qquad \c > 1\,.
\end{equation}
As in the previous subsections,
we replace the $\F^{a,s}_h$ by $\frac{ \c^h \Gamma (h)^2 }{\Gamma (2 h)}{}\, _2F_1(h,h;2 h;\c)$ to obtain
\bal
\nonumber \cg(\c)
&=-\sum_{i \in \{s,a\}} \sum_m \underset{h=h_m^{\bar{i},d}-\frac{1}{2}}{\mathrm{Res}}  \left( \frac{g^{\bar{i}}(h)}{1-k^{o,i}_c(h)}  \frac{ \c^h \Gamma (h)^2 }{\Gamma (2 h)}{}\, _2F_1(h,h;2 h;\c)\right),\quad 0 < \chi < 1.
\eal

This concludes our calculation of the auxiliary quantity $\cg$, but it remains to compute the original
4-point function
\bal
\cf^{\psi b\bar\psi \bar b}(\t_1,\t_2,\t_3,\t_4)=\pa_{\t_4} \cg(\t_1,\t_2,\t_3,\t_4)\,.\label{FdG}
\eal
This $\t_4$ derivative can be worked out explicitly.
Making use of a property of the hypergeometric function and
\bal
\pa_{\t_4}\c=-\frac{\t_{12}\t_{23}}{\t_{13}\t_{24}^2}=-\c \frac{\t_{23}}{\t_{34}\t_{24}}\,,
\eal
the derivative is equivalent to replacing the ${}_2F_1(h,h;2 h;\c)$ by the factor
\bal
 \frac{\mathrm{sgn}(\t_{34})}{|\t_{34}| }\left(\frac{1}{q  } {}\, _2F_1(h,h;2 h;\c)- \frac{\t_{23}}{ \t_{24}}h {}\, _2F_1(h,h+1;2 h;\c) \right). \label{rpl}
\eal
The replacement for ${}\, _2F_1(1-h,1-h;2-2 h;\c)$ can be obtained from~\eqref{rpl} by sending $h\to 1-h$.

Notice that the computation in this subsection is secretly making use of the supersymmetry Ward identity. The supersymmetry transformation on the bosonic fields take the schematic form $Q b \to \pa \psi$. Therefore, the step of taking a derivative in our computation, which we saw essentially converts a fermionic factor into a bosonic one,  is essentially using supersymmetry to relate the computation in this subsection to a similar computation in section~\ref{psi4}.  In fact, the result~\eqref{FdG} has a form that manifestly satisfies a supersymmetry Ward identity.

Furthermore, we note  from the expressions~\eqref{fm} and~\eqref{pb}, that the $\F^s_2(\c)$ term becomes divergent at $h=2=h^{a,-}_0$ and the $\F^a_1(\c)$ term becomes divergent at $h=1=h^{s,+}_0$. From the expression~\eqref{gf}, the $\F^a_1(\c)$  term becomes divergent at $h=1=h^{o,a,+}=h^{s,d}_0$.
This confirms that all divergences arise from contributions from
the first $\cn=2$ supermultiplet
$(h^{s,+}_0,h^{s,d}_0,h^{a,-}_0)$. This agrees with the
expectation of~\cite{Fu:2016vas}.

It would be very interesting
to carefully work out the enhanced contribution from this supermultiplet,
along the lines of the same analysis for the fermionic SYK model~\cite{Maldacena:2016hyu}, but we will not do so here.

\section{6-point Functions}

A natural next step would be to compute 6-point correlation functions to extract the OPE coefficients among the singlet bilinear operators of the model. This would be a supersymmetric generalization of the work~\cite{Gross:2017hcz}, and we will follow many of the notations employed there.
Given all the 4-point functions that we have worked out, we can take their OPE limits and read out the structure constants $c^{xy}_n$ according to
\bal
\frac{1}{N} \sum_{i=1}^N x_i(\t_1)y_i(\t_2)=\text{sgn}(\t_{12})^{{d(x)d(y)}}\frac{|\t_{12}|^{h_n-h_x-h_y}}{\sqrt{N}} \sum_n c^{xy}_n\, F_{\co_n}(\t_2)\,,
\eal
where $x,y \in \{\psi,b\}$, $d(\psi)=1$, $d(b)=0$, and
\bal
F_{\co_n}(\t)= \left(1+\frac{1}{2}\t\pa_{\t}+\cdots\right) \co_n(\t)
\eal
is a family of operators containing the primary $\co_n$ and all of its descendents. In practice, the coefficients $c^{xy}_n$ can be read off from  a 4-point function by restoring all of the time dependence via~\eqref{crossratio}
and expanding it into the form
\bal
\cf^{xyzw}(\t_1,\t_2,\t_3,\t_4)&=\frac{\text{sgn}(\t_{12})^{d(x)d(y)}}{|\t_{12}|^{h_1+h_2}}\frac{\text{sgn}(\t_{34})^{d(z)d(w)}}{|\t_{34}|^{h_3+h_4}}\sum_n \frac{|c^{\f\f}_n|^2 }{|\t_{12}|^{-h_n}|\t_{34}|^{-h_n}|\t_{24}|^{2h_n}}
\eal
in the OPE limit
\bal
|\t_{12}|\ll |\t_{24}|\,,\quad |\t_{34}|\ll |\t_{24}|\,.
\eal
With these coefficients extracted out, one can easily check that there are again two different types of contributing diagrams to the leading order in the large-$N$ limit of the supersymmetric SYK model. Because we are treating the supersymmetric model in component form, the computations are almost the same as those in section 3 of~\cite{Gross:2017hcz}. In particular, we would need to compute exactly the same integrals $I^{(1)}_{mnp}$ and $I^{(2)}_{mnp}$. The only difference is that we have more possible external configurations and thus would need to sum over more combinations of  $c^{\f\f}_n$. Since the computation will be largely identical to those in~\cite{Gross:2017hcz}, we will not elaborate any details here.

\acknowledgments

We thank Micha Berkooz, Antal Jevicki and Moshe Rozali for useful discussions on related topics.
This work was supported by the US Department of Energy under contract DE-SC0010010 Task A and by Simons Investigator Award \#376208 (AV).
CP thanks the Galileo Galilei Institute for Theoretical Physics (GGI) for hospitality within the program ``New Developments in AdS3/CFT2 Holography", during which period he was partially supported by a Young Investigator Training Program fellowship from INFN as well as the ACRI (Associazione di Fondazioni e di Casse di Risparmio S.p.a.).

\appendix

\section{Useful Integrals}\label{evss}

Here we tabulate some useful integrals:
\begin{multline}
 \int dt dt' \frac{\text{sgn}(t_1-t)\text{sgn}(t_2-t')\text{sgn}(t-t')}{|t_1-t|^{2\a}|t_2-t'|^{2\b}|t-t'|^{2\g}}\\
= \frac{\p^2   \sin(\p(\a+\b+\g-1)\Gamma(2\a+2\b+2\g-2)}{  \sin(\p\a)\Gamma(2\a) \sin(\p\b)\Gamma(2\b)\sin(\p\g)\Gamma(2\g)}    \frac{\text{sgn}(t_{12})}{|t_{12}|^{ 2\a+2\b+2\g-2}}\,,\label{sss}
\end{multline}
\begin{multline}
 \int dt dt' \frac{\text{sgn}(t_1-t)\text{sgn}(t_2-t')}{|t_1-t|^{2\a}|t_2-t'|^{2\b}|t-t'|^{2\g}}\\
 = \frac{\p^2   \cos(\p(\a+\b+\g-1)\Gamma(2\a+2\b+2\g-2)}{  \sin(\p\a)\Gamma(2\a) \sin(\p\b)\Gamma(2\b)\cos(\p\g)\Gamma(2\g)}    \frac{1}{|t_{12}|^{ 2\a+2\b+2\g-2}}\,,\label{ss1}
 \end{multline}
\begin{multline}
 \int dt dt' \frac{\text{sgn}(t_1-t) \text{sgn}(t-t')}{|t_1-t|^{2\a}|t_2-t'|^{2\b}|t-t'|^{2\g}}\\
 =- \frac{\p^2   \cos(\p(\a+\b+\g-1)\Gamma(2\a+2\b+2\g-2)}{  \sin(\p\a)\Gamma(2\a) \cos(\p\b)\Gamma(2\b)\sin(\p\g)\Gamma(2\g)}    \frac{1}{|t_{12}|^{ 2\a+2\b+2\g-2}}\,,\label{s1s}
\end{multline}\begin{multline}
\int dt dt' \frac{\text{sgn}(t_1-t) }{|t_1-t|^{2\a}|t_2-t'|^{2\b}|t-t'|^{2\g}}\\
=  \frac{\p^2   \sin(\p(\a+\b+\g-1)\Gamma(2\a+2\b+2\g-2)}{  \sin(\p\a)\Gamma(2\a) \cos(\p\b)\Gamma(2\b)\cos(\p\g)\Gamma(2\g)}    \frac{\text{sgn}(t_{12})}{|t_{12}|^{ 2\a+2\b+2\g-2}}\,.\label{s11}
\end{multline}
\begin{multline}
 \int dt dt' \frac{  \text{sgn}(t-t')}{|t_1-t|^{2\a}|t_2-t'|^{2\b}|t-t'|^{2\g}}\\
  = \frac{\p^2   \sin(\p(\a+\b+\g-1)\Gamma(2\a+2\b+2\g-2)}{  \cos(\p\a)\Gamma(2\a) \cos(\p\b)\Gamma(2\b)\sin(\p\g)\Gamma(2\g)}    \frac{\text{sgn}(t_{12})}{|t_{12}|^{ 2\a+2\b+2\g-2}}\,,\label{11s}
\end{multline}
and
\begin{multline}
 \int dt dt' \frac{ 1 }{|t_1-t|^{2\a}|t_2-t'|^{2\b}|t-t'|^{2\g}}\\
 =  \frac{\p^2   \cos(\p(\a+\b+\g-1)\Gamma(2\a+2\b+2\g-2)}{  \cos(\p\a)\Gamma(2\a) \cos(\p\b)\Gamma(2\b)\cos(\p\g)\Gamma(2\g)}    \frac{1}{|t_{12}|^{ 2\a+2\b+2\g-2}}\,.\label{111}
\end{multline}
Additional similar types of integrals may be found in section~3 of~\cite{Peng:2017kro}. Notice that the integral~\eqref{ss1}, \eqref{s1s} and~\eqref{111} are proportional to $\delta(\t_{12})$ when $2\a+2\b+2\g=3$, similar to those listed in~\cite{Peng:2017kro}.  We omitted these special cases in the above table for simplicity.

These results can be derived from slightly tedious computations building on
the basic identities~\cite{Fu:2016vas}
\bal
\frac{1}{|t|^{2 \Delta}} &= \frac{1}{2 \cos(\pi \Delta) \Gamma(2 \Delta)} \int d\omega\, e^{i \omega t} \frac{1}{|w|^{1-2 \Delta}}\,,\\
\frac{\text{sgn}(t)}{|t|^{2\Delta}}& =\frac{1}{2i \sin(\p\Delta)\Gamma(2\Delta)} \int dw \, e^{i w t} \frac{\text{sgn}(w)}{|w|^{1-2\Delta}}\,.
\eal
Here we provide one example of such a derivation:
 \bal
 &\int dt dt' \frac{\text{sgn}(t_1-t)\text{sgn}(t_2-t')\text{sgn}(t-t')}{|t_1-t|^{2\a}|t_2-t'|^{2\b}|t-t'|^{2\g}}\\
 &=\int dt dt' \frac{1}{2i \sin(\p\a)\Gamma(2\a)} \frac{1}{2i \sin(\p\b)\Gamma(2\b)} \frac{1}{2i \sin(\p\g)\Gamma(2\g)} \\
 &\times \int dw \, e^{i w (t_1-t)} \frac{\text{sgn}(w)}{|w|^{1-2\a}} \int du \, e^{i u (t_2-t')} \frac{\text{sgn}(u)}{|u|^{1-2\b}} \int dv \, e^{i v (t-t')} \frac{\text{sgn}(v)}{|v|^{1-2\g}}\\
  &= \frac{(2\p)}{2i \sin(\p\a)\Gamma(2\a)} \frac{1}{2i \sin(\p\b)\Gamma(2\b)} \frac{1}{2i \sin(\p\g)\Gamma(2\g)} \\
 &\times \int dw \int dt \, e^{i \big(w t_1+u t_2+(-u- w) t \big)} \frac{\text{sgn}(w)}{|w|^{1-2\a}} \int du  \frac{\text{sgn}(u)}{|u|^{1-2\b}}  \frac{\text{sgn}(-u)}{|u|^{1-2\g}}\\
  &= \frac{(2\p)^2}{2i \sin(\p\a)\Gamma(2\a)} \frac{1}{2i \sin(\p\b)\Gamma(2\b)} \frac{1}{2i \sin(\p\g)\Gamma(2\g)} \\
 &\times \int dw  \, e^{i \big(w (t_1- t_2)  \big)} \frac{\text{sgn}(w)}{|w|^{1-2\a}}   \frac{\text{sgn}(-w)}{|w|^{1-2\b}}  \frac{\text{sgn}(w)}{|w|^{1-2\g}}\\
  &= \frac{\p^2   \sin(\p(\a+\b+\g-1)\Gamma(2\a+2\b+2\g-2)}{  \sin(\p\a)\Gamma(2\a) \sin(\p\b)\Gamma(2\b)\sin(\p\g)\Gamma(2\g)}    \frac{\text{sgn}(t_{12})}{|t_{12}|^{ 2\a+2\b+2\g-2}}\,.
 \eal


\begin{thebibliography}{99}

\bibitem{Sachdev:1992fk}
  S.~Sachdev and J.~Ye,
  Phys.\ Rev.\ Lett.\  {\bf 70}, 3339 (1993)
  doi:10.1103/PhysRevLett.70.3339
  [cond-mat/9212030].

\bibitem{PG}
  O.~Parcollet and A.~Georges,
  Phys.\ Rev.\  B {\bf  59}, 5341 (1999)
  doi: 10.1103/PhysRevB.59.5341
  [arXiv:cond-mat/9806119].

\bibitem{GPS}
  A.~Georges, O.~Parcollet, and S.~Sachdev,
  Phys.\ Rev.\ Lett.\   {\bf 85}, 840 (2000)
  [arXiv:cond-mat/9909239].

\bibitem{KitaevTalk1}
  A.~Kitaev,
  \url{http://online.kitp.ucsb.edu/online/joint98/kitaev/}\\

\bibitem{KitaevTalk2}
  A.~Kitaev,
  \url{http://online.kitp.ucsb.edu/online/entangled15/kitaev/}
  \url{http://online.kitp.ucsb.edu/online/entangled15/kitaev2/}

\bibitem{Maldacena:2016hyu}
  J.~Maldacena and D.~Stanford,
  Phys.\ Rev.\ D {\bf 94}, no. 10, 106002 (2016)
  doi:10.1103/PhysRevD.94.106002
  [arXiv:1604.07818 [hep-th]].

\bibitem{Strominger:1998yg}
  A.~Strominger,
  JHEP {\bf 9901}, 007 (1999)
  doi:10.1088/1126-6708/1999/01/007
  [hep-th/9809027].

\bibitem{Maldacena:1998uz}
  J.~M.~Maldacena, J.~Michelson and A.~Strominger,
  JHEP {\bf 9902}, 011 (1999)
  doi:10.1088/1126-6708/1999/02/011
  [hep-th/9812073].

\bibitem{Sachdev:2010um}
  S.~Sachdev,
  Phys.\ Rev.\ Lett.\  {\bf 105}, 151602 (2010)
  doi:10.1103/PhysRevLett.105.151602
  [arXiv:1006.3794 [hep-th]].

\bibitem{Shenker:2013pqa}
  S.~H.~Shenker and D.~Stanford,
  JHEP {\bf 1403}, 067 (2014)
  doi:10.1007/JHEP03(2014)067
  [arXiv:1306.0622 [hep-th]].

\bibitem{Almheiri:2014cka}
  A.~Almheiri and J.~Polchinski,
  JHEP {\bf 1511}, 014 (2015)
  doi:10.1007/JHEP11(2015)014
  [arXiv:1402.6334 [hep-th]].

\bibitem{Shenker:2014cwa}
  S.~H.~Shenker and D.~Stanford,
  JHEP {\bf 1505}, 132 (2015)
  doi:10.1007/JHEP05(2015)132
  [arXiv:1412.6087 [hep-th]].

\bibitem{Maldacena:2015waa}
  J.~Maldacena, S.~H.~Shenker and D.~Stanford,
  JHEP {\bf 1608}, 106 (2016)
  doi:10.1007/JHEP08(2016)106
  [arXiv:1503.01409 [hep-th]].

\bibitem{Sachdev:2015efa}
  S.~Sachdev,
  Phys.\ Rev.\ X {\bf 5}, no. 4, 041025 (2015)
  doi:10.1103/PhysRevX.5.041025
  [arXiv:1506.05111 [hep-th]].

\bibitem{Blake:2016jnn}
  M.~Blake and A.~Donos,
  JHEP {\bf 1702}, 013 (2017)
  doi:10.1007/JHEP02(2017)013
  [arXiv:1611.09380 [hep-th]].

\bibitem{Maldacena:2017axo}
  J.~Maldacena, D.~Stanford and Z.~Yang,
  Fortsch.\ Phys.\  {\bf 65}, no. 5, 1700034 (2017)
  doi:10.1002/prop.201700034
  [arXiv:1704.05333 [hep-th]].

\bibitem{Bagrets:2016cdf}
  D.~Bagrets, A.~Altland and A.~Kamenev,
  Nucl.\ Phys.\ B {\bf 911}, 191 (2016)
  doi:10.1016/j.nuclphysb.2016.08.002
  [arXiv:1607.00694 [cond-mat.str-el]].

\bibitem{Stanford:2017thb}
  D.~Stanford and E.~Witten,
  arXiv:1703.04612 [hep-th].

\bibitem{Mertens:2017mtv}
  T.~G.~Mertens, G.~J.~Turiaci and H.~L.~Verlinde,
  arXiv:1705.08408 [hep-th].

\bibitem{Maldacena:2016upp}
  J.~Maldacena, D.~Stanford and Z.~Yang,
  PTEP {\bf 2016}, no. 12, 12C104 (2016)
  doi:10.1093/ptep/ptw124
  [arXiv:1606.01857 [hep-th]].

\bibitem{Engelsoy:2016xyb}
  J.~Engels\"{o}y, T.~G.~Mertens and H.~Verlinde,
  JHEP {\bf 1607}, 139 (2016)
  doi:10.1007/JHEP07(2016)139
  [arXiv:1606.03438 [hep-th]].

\bibitem{Cvetic:2016eiv}
  M.~Cveti\v{c} and I.~Papadimitriou,
  JHEP {\bf 1612}, 008 (2016)
  Erratum: [JHEP {\bf 1701}, 120 (2017)]
  doi:10.1007/JHEP12(2016)008, 10.1007/JHEP01(2017)120
  [arXiv:1608.07018 [hep-th]].

\bibitem{Forste:2017kwy}
  S.~Forste and I.~Golla,
  Phys.\ Lett.\ B {\bf 771}, 157 (2017)
  doi:10.1016/j.physletb.2017.05.039
  [arXiv:1703.10969 [hep-th]].

\bibitem{Polchinski:2016xgd}
  J.~Polchinski and V.~Rosenhaus,
  JHEP {\bf 1604}, 001 (2016)
  doi:10.1007/JHEP04(2016)001
  [arXiv:1601.06768 [hep-th]].

\bibitem{Jevicki:2016bwu}
  A.~Jevicki, K.~Suzuki and J.~Yoon,
  JHEP {\bf 1607}, 007 (2016)
  doi:10.1007/JHEP07(2016)007
  [arXiv:1603.06246 [hep-th]].

\bibitem{Jevicki:2016ito}
  A.~Jevicki and K.~Suzuki,
  JHEP {\bf 1611}, 046 (2016)
  doi:10.1007/JHEP11(2016)046
  [arXiv:1608.07567 [hep-th]].

\bibitem{Garcia-Garcia:2016mno}
  A.~M.~Garc\'{i}a-Garc\'{i}a and J.~J.~M.~Verbaarschot,
  Phys.\ Rev.\ D {\bf 94}, no. 12, 126010 (2016)
  doi:10.1103/PhysRevD.94.126010
  [arXiv:1610.03816 [hep-th]].

\bibitem{Cotler:2016fpe}
  J.~S.~Cotler {\it et al.},
  JHEP {\bf 1705}, 118 (2017)
  doi:10.1007/JHEP05(2017)118
  [arXiv:1611.04650 [hep-th]].

\bibitem{Liu:2016rdi}
  Y.~Liu, M.~A.~Nowak and I.~Zahed,
  arXiv:1612.05233 [hep-th].

\bibitem{Garcia-Garcia:2017pzl}
  A.~M.~Garc\'{i}a-Garc\'{i}a and J.~J.~M.~Verbaarschot,
  arXiv:1701.06593 [hep-th].

\bibitem{Li:2017hdt}
  T.~Li, J.~Liu, Y.~Xin and Y.~Zhou,
  JHEP {\bf 1706}, 111 (2017)
  doi:10.1007/JHEP06(2017)111
  [arXiv:1702.01738 [hep-th]].

\bibitem{Gross:2017hcz}
  D.~J.~Gross and V.~Rosenhaus,
  JHEP {\bf 1705}, 092 (2017)
  doi:10.1007/JHEP05(2017)092
  [arXiv:1702.08016 [hep-th]].

\bibitem{Gurau:2011xq}
  R.~Gurau,
  Annales Henri Poincare {\bf 13}, 399 (2012)
  doi:10.1007/s00023-011-0118-z
  [arXiv:1102.5759 [gr-qc]].

\bibitem{Bonzom:2011zz}
  V.~Bonzom, R.~Gurau, A.~Riello and V.~Rivasseau,
  Nucl.\ Phys.\ B {\bf 853}, 174 (2011)
  doi:10.1016/j.nuclphysb.2011.07.022
  [arXiv:1105.3122 [hep-th]].

\bibitem{Bonzom:2012hw}
  V.~Bonzom, R.~Gurau and V.~Rivasseau,
  Phys.\ Rev.\ D {\bf 85}, 084037 (2012)
  doi:10.1103/PhysRevD.85.084037
  [arXiv:1202.3637 [hep-th]].

\bibitem{Carrozza:2015adg}
  S.~Carrozza and A.~Tanasa,
  Lett.\ Math.\ Phys.\  {\bf 106}, no. 11, 1531 (2016)
  doi:10.1007/s11005-016-0879-x
  [arXiv:1512.06718 [math-ph]].

\bibitem{Witten:2016iux}
  E.~Witten,
  arXiv:1610.09758 [hep-th].

\bibitem{Gurau:2016lzk}
  R.~Gurau,
  Nucl.\ Phys.\ B {\bf 916}, 386 (2017)
  doi:10.1016/j.nuclphysb.2017.01.015
  [arXiv:1611.04032 [hep-th]].

\bibitem{Klebanov:2016xxf}
  I.~R.~Klebanov and G.~Tarnopolsky,
  Phys.\ Rev.\ D {\bf 95}, no. 4, 046004 (2017)
  doi:10.1103/PhysRevD.95.046004
  [arXiv:1611.08915 [hep-th]].

\bibitem{Nishinaka:2016nxg}
  T.~Nishinaka and S.~Terashima,
  arXiv:1611.10290 [hep-th].

\bibitem{Peng:2016mxj}
  C.~Peng, M.~Spradlin and A.~Volovich,
  JHEP {\bf 1705}, 062 (2017)
  doi:10.1007/JHEP05(2017)062
  [arXiv:1612.03851 [hep-th]].

\bibitem{Krishnan:2016bvg}
  C.~Krishnan, S.~Sanyal and P.~N.~Bala Subramanian,
  JHEP {\bf 1703}, 056 (2017)
  doi:10.1007/JHEP03(2017)056
  [arXiv:1612.06330 [hep-th]].

\bibitem{Gurau:2017xhf}
  R.~Gurau,
  arXiv:1702.04228 [hep-th].

\bibitem{Bonzom:2017pqs}
  V.~Bonzom, L.~Lionni and A.~Tanasa,
  J.\ Math.\ Phys.\  {\bf 58}, no. 5, 052301 (2017)
  doi:10.1063/1.4983562
  [arXiv:1702.06944 [hep-th]].

\bibitem{Krishnan:2017ztz}
  C.~Krishnan, K.~V.~P.~Kumar and S.~Sanyal,
  JHEP {\bf 1706}, 036 (2017)
  doi:10.1007/JHEP06(2017)036
  [arXiv:1703.08155 [hep-th]].

\bibitem{Narayan:2017qtw}
  P.~Narayan and J.~Yoon,
  arXiv:1705.01554 [hep-th].

\bibitem{Klebanov:2017nlk}
  I.~R.~Klebanov and G.~Tarnopolsky,
  arXiv:1706.00839 [hep-th].

\bibitem{Mironov:2017aqv}
  A.~Mironov and A.~Morozov,
  arXiv:1706.03667 [hep-th].

\bibitem{Peng:2017kro}
  C.~Peng,
  JHEP {\bf 1705}, 129 (2017)
  doi:10.1007/JHEP05(2017)129
  [arXiv:1704.04223 [hep-th]].

\bibitem{Gross:2016kjj}
  D.~J.~Gross and V.~Rosenhaus,
  JHEP {\bf 1702}, 093 (2017)
  doi:10.1007/JHEP02(2017)093
  [arXiv:1610.01569 [hep-th]].

\bibitem{Gu:2016oyy}
  Y.~Gu, X.~L.~Qi and D.~Stanford,
  JHEP {\bf 1705}, 125 (2017)
  doi:10.1007/JHEP05(2017)125
  [arXiv:1609.07832 [hep-th]].

\bibitem{Berkooz:2016cvq}
  M.~Berkooz, P.~Narayan, M.~Rozali and J.~Sim\'{o}n,
  JHEP {\bf 1701}, 138 (2017)
  doi:10.1007/JHEP01(2017)138
  [arXiv:1610.02422 [hep-th]].

\bibitem{Turiaci:2017zwd}
  G.~Turiaci and H.~Verlinde,
  arXiv:1701.00528 [hep-th].

\bibitem{Gu:2017ohj}
  Y.~Gu, A.~Lucas and X.~L.~Qi,
  SciPost Phys.\  {\bf 2}, no. 3, 018 (2017)
  doi:10.21468/SciPostPhys.2.3.018
  [arXiv:1702.08462 [hep-th]].

\bibitem{Jian:2017unn}
  S.~K.~Jian and H.~Yao,
  arXiv:1703.02051 [cond-mat.str-el].

\bibitem{Das:2017pif}
  S.~R.~Das, A.~Jevicki and K.~Suzuki,
  arXiv:1704.07208 [hep-th].

\bibitem{You:2016ldz}
  Y.~Z.~You, A.~W.~W.~Ludwig and C.~Xu,
  Phys.\ Rev.\ B {\bf 95}, no. 11, 115150 (2017)
  doi:10.1103/PhysRevB.95.115150
  [arXiv:1602.06964 [cond-mat.str-el]].

\bibitem{Fu:2016yrv}
  W.~Fu and S.~Sachdev,
  Phys.\ Rev.\ B {\bf 94}, no. 3, 035135 (2016)
  doi:10.1103/PhysRevB.94.035135
  [arXiv:1603.05246 [cond-mat.str-el]].

\bibitem{Danshita:2016xbo}
  I.~Danshita, M.~Hanada and M.~Tezuka,
  arXiv:1606.02454 [cond-mat.quant-gas].

\bibitem{Garcia-Alvarez:2016wem}
  L.~Garc\'{i}a-\'{A}lvarez, I.~L.~Egusquiza, L.~Lamata, A.~del Campo, J.~Sonner and E.~Solano,
  Phys.\ Rev.\ Lett.\  {\bf 119}, 040501 (2017)
  doi:10.1103/PhysRevLett.119.040501
  [arXiv:1607.08560 [quant-ph]].

\bibitem{Banerjee:2016ncu}
  S.~Banerjee and E.~Altman,
  Phys.\ Rev.\ B {\bf 95}, no. 13, 134302 (2017)
  doi:10.1103/PhysRevB.95.134302
  [arXiv:1610.04619 [cond-mat.str-el]].

\bibitem{Davison:2016ngz}
  R.~A.~Davison, W.~Fu, A.~Georges, Y.~Gu, K.~Jensen and S.~Sachdev,
  Phys.\ Rev.\ B {\bf 95}, no. 15, 155131 (2017)
  doi:10.1103/PhysRevB.95.155131
  [arXiv:1612.00849 [cond-mat.str-el]].

\bibitem{Hartnoll:2016apf}
  S.~A.~Hartnoll, A.~Lucas and S.~Sachdev,
  arXiv:1612.07324 [hep-th].

\bibitem{Bi:2017yvx}
  Z.~Bi, C.~M.~Jian, Y.~Z.~You, K.~A.~Pawlak and C.~Xu,
  Phys.\ Rev.\ B {\bf 95}, no. 20, 205105 (2017)
  doi:10.1103/PhysRevB.95.205105
  [arXiv:1701.07081 [cond-mat.str-el]].

\bibitem{Jian:2017jfl}
  C.~M.~Jian, Z.~Bi and C.~Xu,
  arXiv:1703.07793 [cond-mat.str-el].

\bibitem{Song:2017pfw}
  X.~Y.~Song, C.~M.~Jian and L.~Balents,
  arXiv:1705.00117 [cond-mat.str-el].

\bibitem{Wu:2017exh}
  S.~F.~Wu, B.~Wang, X.~H.~Ge and Y.~Tian,
  arXiv:1706.00718 [hep-th].

\bibitem{Jensen:2016pah}
  K.~Jensen,
  Phys.\ Rev.\ Lett.\  {\bf 117}, no. 11, 111601 (2016)
  doi:10.1103/PhysRevLett.117.111601
  [arXiv:1605.06098 [hep-th]].

\bibitem{Ferrari:2017ryl}
  F.~Ferrari,
  arXiv:1701.01171 [hep-th].

\bibitem{Ho:2017nyc}
  W.~W.~Ho and D.~Radicevic,
  arXiv:1701.08777 [quant-ph].

\bibitem{Berkooz:2017efq}
  M.~Berkooz, P.~Narayan, M.~Rozali and J.~Sim\'{o}n,
  arXiv:1702.05105 [hep-th].

\bibitem{Cotler:2017abq}
  J.~S.~Cotler, G.~R.~Penington and D.~H.~Ranard,
  arXiv:1702.06142 [quant-ph].

\bibitem{Bagrets:2017pwq}
  D.~Bagrets, A.~Altland and A.~Kamenev,
  Nucl.\ Phys.\ B {\bf 921}, 727 (2017)
  doi:10.1016/j.nuclphysb.2017.06.012
  [arXiv:1702.08902 [cond-mat.str-el]].

\bibitem{Caputa:2017urj}
  P.~Caputa, N.~Kundu, M.~Miyaji, T.~Takayanagi and K.~Watanabe,
  arXiv:1703.00456 [hep-th].

\bibitem{Chowdhury:2017jzb}
  D.~Chowdhury and B.~Swingle,
  arXiv:1703.02545 [cond-mat.str-el].

\bibitem{Itoyama:2017emp}
  H.~Itoyama, A.~Mironov and A.~Morozov,
  Phys.\ Lett.\ B {\bf 771}, 180 (2017)
  doi:10.1016/j.physletb.2017.05.043
  [arXiv:1703.04983 [hep-th]].

\bibitem{Itoyama:2017xid}
  H.~Itoyama, A.~Mironov and A.~Morozov,
  JHEP {\bf 1706}, 115 (2017)
  doi:10.1007/JHEP06(2017)115
  [arXiv:1704.08648 [hep-th]].

\bibitem{Blake:2017qgd}
  M.~Blake, R.~A.~Davison and S.~Sachdev,
  arXiv:1705.07896 [hep-th].

\bibitem{Gurau:2017qna}
  R.~Gurau,
  arXiv:1705.08581 [hep-th].

\bibitem{Dartois:2017xoe}
  S.~Dartois, H.~Erbin and S.~Mondal,
  arXiv:1706.00412 [hep-th].

\bibitem{Kanazawa:2017dpd}
  T.~Kanazawa and T.~Wettig,
  arXiv:1706.03044 [hep-th].

\bibitem{Gurau:2017qya}
  R.~Gurau,
  arXiv:1706.05328 [hep-th].

\bibitem{Krishnan:2017txw}
  C.~Krishnan and K.~V.~P.~Kumar,
  arXiv:1706.05364 [hep-th].

\bibitem{Fu:2016vas}
  W.~Fu, D.~Gaiotto, J.~Maldacena and S.~Sachdev,
  Phys.\ Rev.\ D {\bf 95}, no. 2, 026009 (2017)
  Addendum: [Phys.\ Rev.\ D {\bf 95}, no. 6, 069904 (2017)]
  doi:10.1103/PhysRevD.95.069904, 10.1103/PhysRevD.95.026009
  [arXiv:1610.08917 [hep-th]].

\bibitem{Anninos:2016szt}
  D.~Anninos, T.~Anous and F.~Denef,
  JHEP {\bf 1612}, 071 (2016)
  doi:10.1007/JHEP12(2016)071
  [arXiv:1603.00453 [hep-th]].

\bibitem{Sannomiya:2016mnj}
  N.~Sannomiya, H.~Katsura and Y.~Nakayama,
  Phys.\ Rev.\ D {\bf 95}, no. 6, 065001 (2017)
  doi:10.1103/PhysRevD.95.065001
  [arXiv:1612.02285 [cond-mat.str-el]].

\bibitem{Murugan:2017eto}
  J.~Murugan, D.~Stanford and E.~Witten,
  arXiv:1706.05362 [hep-th].

\bibitem{Yoon:2017gut}
  J.~Yoon,
  arXiv:1706.05914 [hep-th].

\bibitem{Bulycheva:2017uqj}
  K.~Bulycheva,
  arXiv:1706.07411 [hep-th].

\end{thebibliography}
\end{document}